\documentclass[sigconf, authorversion=true]{acmart}

\newcommand{\tool}{Data Formulator 2\xspace}
\newcommand{\df}{\textsc{Df2}\xspace}

\AtBeginDocument{%
  }

\copyrightyear{2025} 
\acmYear{2025} 
\setcopyright{cc}
\setcctype{by}
\acmConference[CHI '25]{CHI Conference on Human Factors in Computing Systems}{April 26-May 1, 2025}{Yokohama, Japan}
\acmBooktitle{CHI Conference on Human Factors in Computing Systems (CHI '25), April 26-May 1, 2025, Yokohama, Japan}\acmDOI{10.1145/3706598.3713296}
\acmISBN{979-8-4007-1394-1/25/04}

\graphicspath{{figs/}{figures/}{pictures/}{images/}{./}} 

\usepackage{acmart-taps}
\usepackage{enumitem}
\usepackage{xspace}
\usepackage{colortbl}

\usepackage{listings}

\newcommand{\bpstart}[1]{\smallskip\noindent{\textbf{#1.}}}
\newcommand{\code}[1]{{\fontfamily{phv}\selectfont\footnotesize {#1}}}

\usepackage{soul}
\newcommand{\hlc}[2][yellow]{{%
                  \colorlet{foo}{#1}%
                  \sethlcolor{foo}\hl{#2}}%
}
\definecolor{periwinkle}{RGB}{240, 240, 255} 
\definecolor{urlcolor}{RGB}{24,64,127}  
\definecolor{promptcolor}{RGB}{255, 240, 240} 
\newcommand\qt[2]{\hlc[{#1}]{#2}}
\newcommand{\pquote}[1]{\qt{periwinkle}{\emph{#1}}}
\newcommand{\pprompt}[1]{\qt{promptcolor}{\emph{#1}}}

\newsavebox{\fmbox}
\newenvironment{smpage}[1]
{\begin{lrbox}{\fmbox}\begin{minipage}{#1}}
{\end{minipage}\end{lrbox}\usebox{\fmbox}}

\definecolor{tempcolor}{RGB}{44,94,127}  
\newcommand*\circled[1]{%
  \raisebox{.5pt}{\textcircled{\raisebox{-.9pt} {\sf #1}}}%
}  
  
\newcommand*\filled[1]{%
  \begingroup  
  \setlength{\unitlength}{1ex}%
  \begin{picture}(2,2)  
  \put(1.1,0.75){\color{tempcolor}\circle*{2}}  
  \put(1.1,0.75){\makebox(0,0){\textcolor{white}{\sf #1}}}  
  \end{picture}%
  \endgroup  
} 

\lstdefinelanguage{json}{  
  basicstyle=\footnotesize\sffamily,  
  breaklines=true,  
  columns=fullflexible,  
  escapeinside={||},  
  xleftmargin=0pt,  
  xrightmargin=0pt,  
  backgroundcolor=\color{white},  
  showstringspaces=false,  
  frame=none,  
  literate=  
    *{:}{{{\color{tempcolor}:}}}{1}  
     {,}{{{\color{tempcolor},}}}{1}  
     {\{}{{{\color{tempcolor}\{}}}{1}  
     {\}}{{{\color{tempcolor}\}}}}{1}  
     {[}{{{\color{tempcolor}[}}}{1}  
     {]}{{{\color{tempcolor}]}}}{1},  
  keywordstyle=\bf\color{tempcolor},  
  stringstyle=\color{red},  
  commentstyle=\color{gray},  
  morekeywords={detailed_instruction, output_fields, visualization_fields, reason}, 
}

\begin{document}

\title{\tool: Iterative Creation of Data Visualizations, with AI Transforming Data Along the Way}

\author{Chenglong Wang}
\email{chenglong.wang@microsoft.com}
\affiliation{%
  \institution{Microsoft Research}
  \city{Redmond}
  \state{Washington}
  \country{USA}
}

\author{Bongshin Lee}
\email{b.lee@yonsei.ac.kr}
\affiliation{%
  \institution{Yonsei University}
  \city{Seoul}
  \country{Korea}
}

\author{Steven Drucker}
\email{sdrucker@microsoft.com}
\affiliation{%
  \institution{Microsoft Research}
  \city{Redmond}
  \state{Washington}
  \country{USA}
}

\author{Dan Marshall}
\email{danmar@microsoft.com}
\affiliation{%
  \institution{Microsoft Research}
  \city{Redmond}
  \state{Washington}
  \country{USA}
}

\author{Jianfeng Gao}
\email{jfgao@microsoft.com}
\affiliation{%
  \institution{Microsoft Research}
  \city{Redmond}
  \state{Washington}
  \country{USA}
}


\begin{abstract}
Data analysts often need to iterate between data transformations and chart designs to create rich visualizations for exploratory data analysis. Although many AI-powered systems have been introduced to reduce the effort of visualization authoring, existing systems are not well suited for iterative authoring. They typically require analysts to provide, in a single turn, a text-only prompt that fully describe a complex visualization. 
We introduce \tool (\df for short), an AI-powered visualization system designed to overcome this limitation.
\df blends graphical user interfaces and natural language inputs to enable users to convey their intent more effectively, while delegating data transformation to AI.
Furthermore, to support efficient iteration, \df lets users navigate their iteration history and reuse previous designs, eliminating the need to start from scratch each time. 
A user study with eight participants demonstrated that \df allowed participants to develop their own iteration styles to complete challenging data exploration sessions.
\end{abstract}


\begin{teaserfigure}
\centering
  \includegraphics[width=\linewidth]{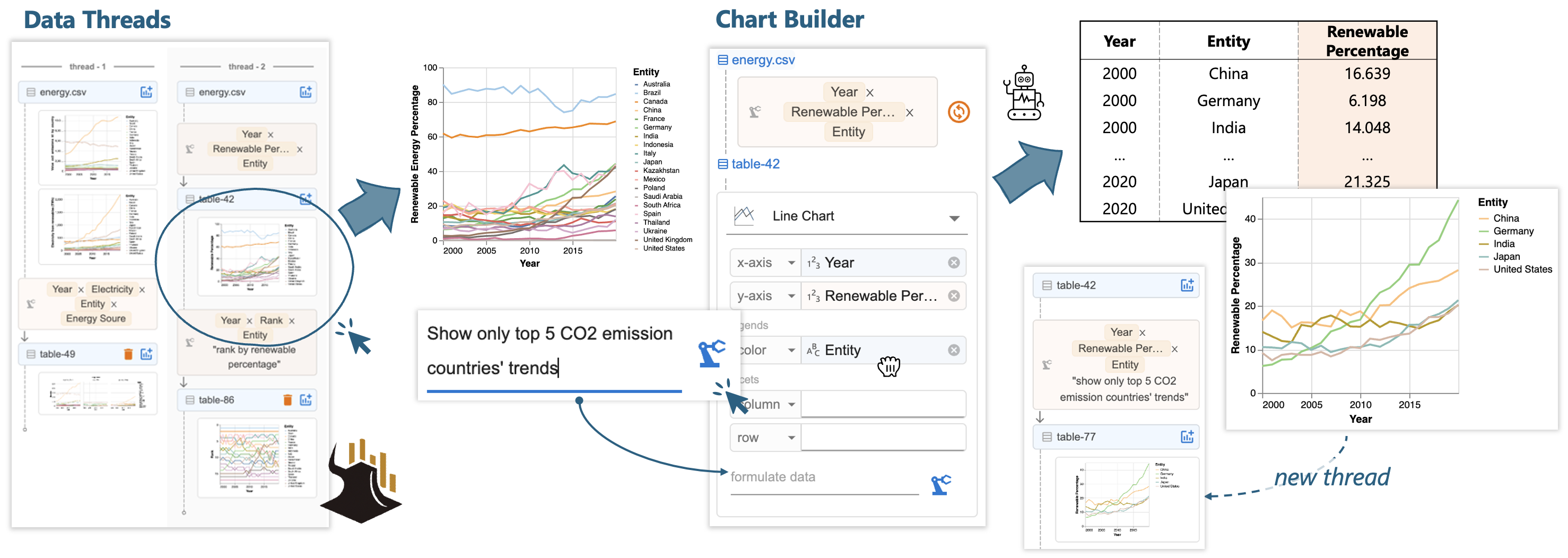}
    \caption{With \tool, analysts can iterate on a previous design by (1) selecting a chart from data threads and (2) providing combined natural language and graphical user interface inputs in the chart builder to specify the new design. The AI model  generates code to transform the data and update the chart. Data threads are updated with new charts for future use.}
    \label{fig:data-anvil-teaser}
\end{teaserfigure}


\maketitle

\section{Introduction}

In data exploration~\cite{rule2018exploration}, even when starting with an initial idea, analysts often need to go back and forth exploring a variety of charts before reaching their goals. Throughout this iterative process, analysts often discover insights that lead them into new directions. However, analysts need to tackle numerous execution challenges: in addition to varying chart specifications (as many current tools facilitate), they need to perform and manage different data transformations to support the desired visualization designs. 
For example, when exploring renewable energy trends, an analyst may find that similar trends across countries make a simple line chart (\autoref{fig:data-anvil-teaser}) too dense for detailed comparisons. This observation prompts the analyst to explore the renewable percentage trends of the top 5 CO$_2$ emitters and how the rankings of these countries have changed over time. To execute the plan, the analyst needs different data transformations: the first requires filtering the data based on each country's total CO$_2$ emissions, and the second requires partitioning the data by year to compute each country's ranking for that year.

Because data transformation can be difficult to learn and execute, many AI-powered tools have been developed~\cite{narechania2020nl4dv,wang2023data,wang2021falx,maddigan2023chat2vis,barke2023grounded,dibia2023lida}. These tools allow users to describe their goals using natural language and leverage AI models' code generation capabilities~\cite{chen2021evaluating,achiam2023gpt} to streamline data transformation and chart creation. Despite their success, current tools do not perform well in the \textit{iterative} visualization authoring context. Most of them require analysts to provide, in a single turn, a {text-only prompt} that {fully describes the complex visualization task to be performed}, which is usually unrealistic for both users and models.
\begin{itemize}[leftmargin=*]
\item First, even though free-form text prompts provide unbounded expressiveness for users to describe their goals, they miss UI interactions' precision and affordances, making it difficult for users to clearly describe complex chart designs.
For example, to fully elaborate a faceted bar chart design, the user needs a verbose prompt to clearly specify visual encodings; without it, AI models often misinterpret the intent and create undesired charts, thus requiring further disambiguation efforts from the user. 
In fact, writing high-quality prompts requires skill and effort. Even with clear goals,  inexperienced users sometimes find it difficult to clearly describe their intent in texts~\cite{zamfirescu2023johnny,DBLP:conf/chi/TankelevitchKSS24}.
\item Second, existing AI-powered tools do not accommodate branching or backtracking, behaviors that commonly occur in the iterative authoring process.
Using single-turn text-to-vis tools iteratively requires users to re-specify their intent from scratch for each new design, even for minor updates. This also increases the likelihood of the AI model failing, as it must solve a complex task in a single attempt. While chat-based tools~\cite{maddigan2023chat2vis,zheng2024opencodeinterpreter,openai2024gpt4technicalreport} support multi-turn interactions by reusing previous outputs, they struggle with branching contexts. Users often find it difficult to clearly specify which previous messages are relevant for the next iteration. With poorly specified contexts, models may struggle at retrieving important information from the lengthy conversation history to complete the task~\cite{liu2024lost,zhang2023tell,hsieh2024ruler}.
\end{itemize}

To address these iterative chart authoring challenges, our first key insight is to design a {\bf multi-modal chart builder} that blends the shelf-configuration UI~\cite{ren2019charticulator,wang2023data} with natural language (NL) input to enhance users' ability to structurally specify their chart designs. Resembling traditional shelf-configuration UIs, the chart builder lets user drag existing fields to corresponding visual channels to specify visual encodings. Additionally, users can type in field names that do not exist in the current data to express their intent for creating a visualization that requires data transformation. Coupled with a brief supplemental NL text that elaborates the design, the user can effectively communicate their goal to AI. 
Since the system can precisely extract chart configuration from the encoding shelf, the user doesn't need a verbose prompt to explicitly explain the design. The AI model then leverages the combined inputs to generate data transformation code to prepare the data required for the chart. 

Our second key insight is to introduce {\bf data threads}  for users to steer iteration directions. Data threads represent user's non-linear authoring history, allowing users to navigate to an earlier result, fork a new branch, and ask AI to create charts based on that context. This reduces users' input overhead by allowing them to specify incremental updates from a previous result (e.g., ``show only top 5 CO$_2$ emission countries' trends'', \autoref{fig:data-anvil-teaser}) rather than re-describing the full chart design from scratch. This design also benefits the AI models: the model can reuse previously generated code for new tasks to avoid repeating past mistakes, and it remains free from distractions caused by irrelevant messages from other threads. Data threads also provide a shortcut for users to backtrack and revise prompts to update recently created charts, allowing them to quickly clarify ambiguous inputs or fix errors made by AI.

\smallskip

Based on these designs, we developed \tool (\df for short), an AI-powered visualization tool for iterative visualization authoring.~\footnote{\tool is open sourced at \textcolor{urlcolor}{\url{https://github.com/microsoft/data-formulator}}}
\df supports diverse charts powered by the Vega-Lite grammar~\cite{satyanarayan2017vegalite}, and the AI model can flexibly transform data  for different designs, supporting operators like reshaping, filtering, aggregation, window functions, and column derivation. Like other AI tools~\cite{dibia2023lida,wang2023data}, \df provides users with panels to view generated data, transformation code and code explanations to inspect AI-generated contents. To understand how our new interaction designs benefit analysts in solving challenging data visualizations tasks, we conducted a user study consisting of eight participants with varying levels of data science expertise. They were asked to reproduce two professional data scientists' analysis sessions to create a total of 16 visualizations, 12 of which require non-trivial data transformations (e.g., rank categories by a criterion and combine low-ranked ones into one category with the label, ``Others''). The study shows that participants can quickly learn to use \df to solve these complex tasks, and the tool's flexibility and expressiveness allow participants to develop their own iteration, verification, and error correction styles to complete the tasks. Our inductive analysis of study sessions reveals interesting patterns of how users' experiences and expectations about the AI system affected their work styles. In summary, our main contributions are as follows:
\begin{itemize}[leftmargin=*]
    \item We designed new interaction approaches, specifically a multi-modal chart builder and a data threads view, to enhance users' ability to specify chart designs and control iteration directions.
    \item We implemented these designs in \df, an AI-powered interactive tool that supports the iterative creation of visualizations requiring data transformations.
    \item We conducted a user study that discovered data analysts' different iteration styles and rich experiences using our new interaction approaches to complete iterative chart authoring tasks. We observed that analysts  developed different styles iterating with the AI to perform data analysis, reflecting their personal experience and expectation with the AI model.
\end{itemize}

\begin{figure*}[t]
\includegraphics[width=\linewidth]{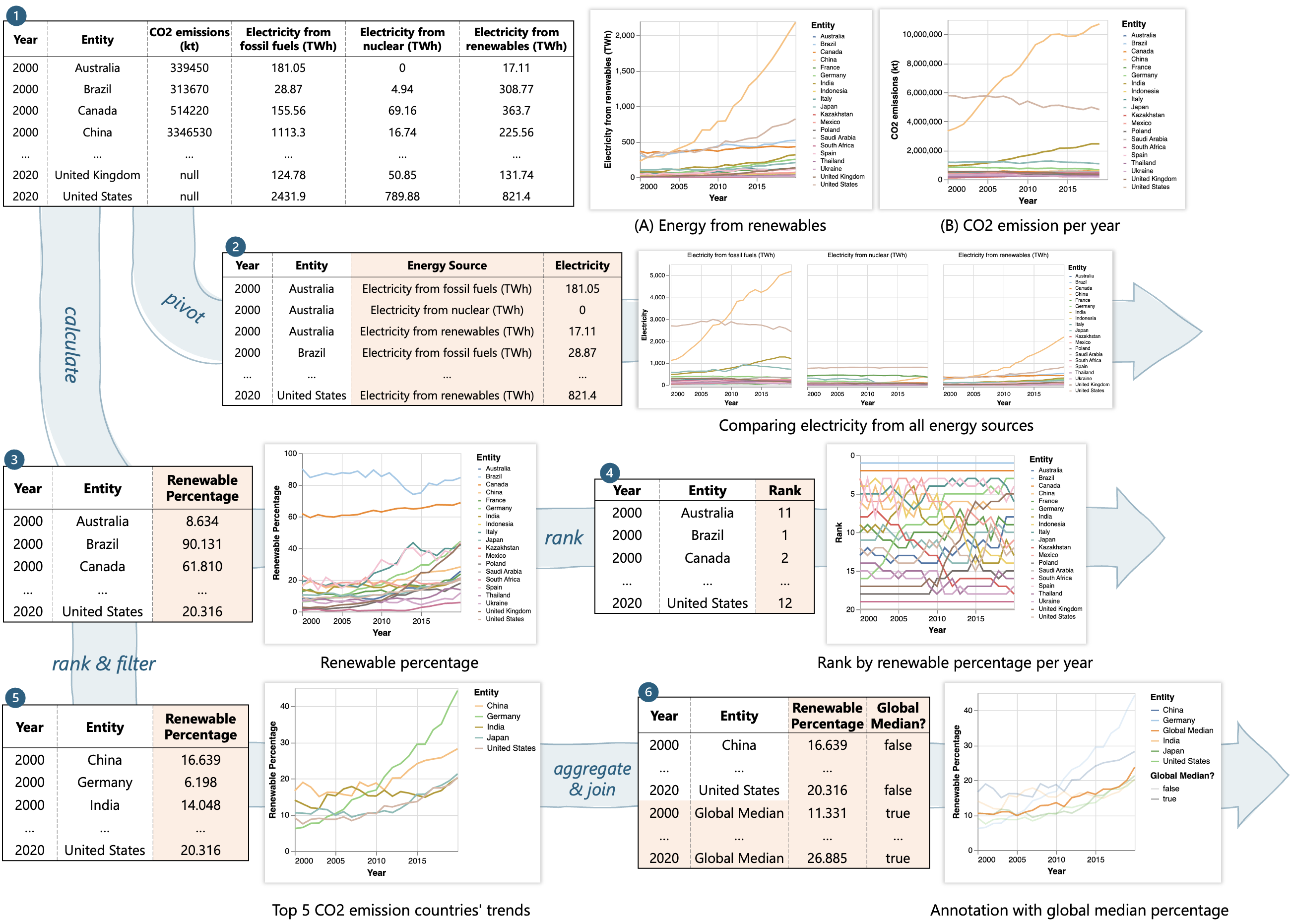}
    \caption{An analyst explores electricity from different energy sources, renewable percentage trends, and country rankings by renewable percentages using a dataset on CO$_2$ and electricity for 20 countries (2000-2020, table 1). The analyst creates five data versions in three branches to support different chart designs. \df allows users to manage iteration directions and create rich visualizations using a blended UI and natural language inputs.}
    \label{fig:example-analysis-session}
\end{figure*}

\section{Illustrative Scenarios}
\label{sec:illustartive-scenarios}

In this section, we describe scenarios to illustrate users’ experiences for creating a series of visualizations to explore global sustainability from a dataset of 20 countries' energy generation from 2000 to 2020. The initial dataset, shown in \autoref{fig:example-analysis-session}-\circled{1}, includes each country's energy produced from three sources (fossil fuel, renewables, and nuclear) each year and annual CO$_2$ emission value (the CO$_2$ emission data only ranges from 2000 to 2019). 
We compare different experiences and skills required for a data analyst, Megan, to complete the analysis session shown in \autoref{fig:example-analysis-session} with different tools, computational notebooks versus \df.

\begin{figure*}[t]
    \centering
    \includegraphics[width=\linewidth]{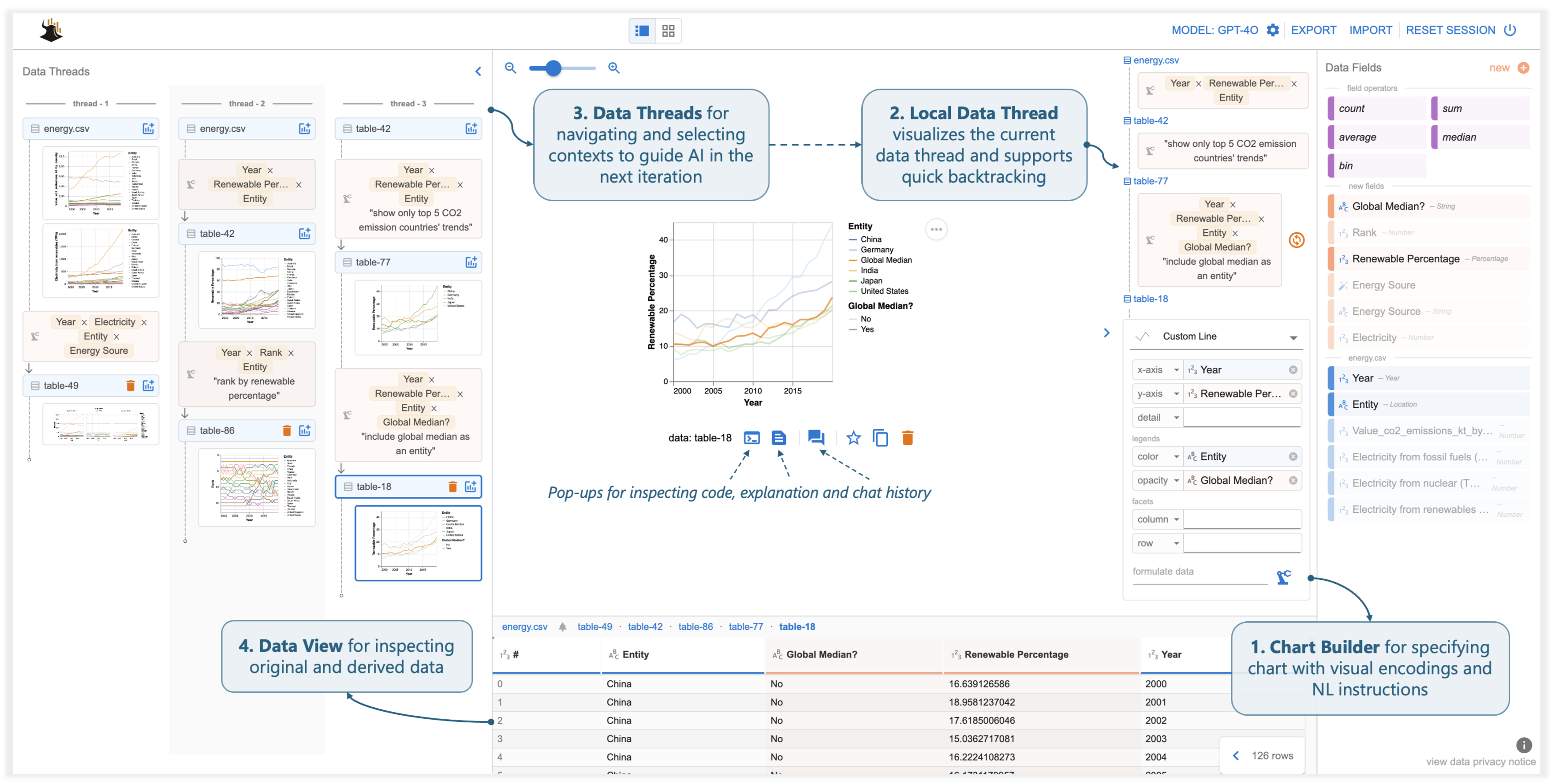}
    \caption{\df overview. Users create visualizations by providing fields (drag-and-drop or type) and NL instructions to the Chart Builder, delegating data transformation to AI. \textbf{Data View} shows derived data. Users navigate data history and select contexts for the next iteration using (the thread in use is displayed as \textbf{local data threads}). They refine or create new charts by providing instructions in \textbf{Chart Builder}. The main panel provides pop-up windows to inspect code, explanations, and chat history.}
    \label{fig:data-anvil-overview}
\end{figure*}

\bpstart{Exploration with computational notebooks}
To complete the analysis in a computation notebook, Megan can use R libraries \textsf{ggplot2} and \textsf{tidyverse}. To use \textsf{ggplot2} to create charts, Megan needs to make sure that all data fields to be visualized on visual channels (e.g., $x,y$-axes, color, facet) are columns in the input data, thus, Megan uses \textsf{tidyverse} to transform data when needed.

\autoref{fig:example-analysis-session} shows Megan's data analysis session with three branches. She starts with two basic line charts (chart \circled{1}-A,B) showing renewable energy and CO2 emission trends. Megan observes that many countries' CO2 emissions have increased despite increased renewable energy use, prompting her to create a faceted line chart (chart \circled{2}) and visualize renewable energy percentage trends (chart \circled{3}). Discovering that renewable percentage is a better indicator for global sustainability trends, Megan explores two directions: creating a line chart of countries' renewable percentage ranks (chart \circled{4}) and highlighting the top 5 CO2 emitters' trends (chart \circled{5}) compared to global median values (chart \circled{6}). Throughout the process, Megan backtracks several times to fork new branches from a previous version of data (e.g., charts \circled{2} to \circled{3}, and \circled{4} to \circled{5}) and reuses existing results to create new charts (e.g., chart \circled{6} from \circled{5}).

Implementing these charts requires considerable data preparation efforts. While basic charts can be created by mapping existing data fields to visual channels (e.g., \code{Year}$\rightarrow x$, \code{Electricity from renewables (Twh)}$\rightarrow y$, \code{Entity}$\rightarrow$\code{color} for chart \circled{1}-A), more complex charts (\circled{3}-\circled{6}) require different data transformations. For example, Megan needs to reshape the table with \code{pivot\_longer} to merge energy sources into a new field \code{Electricity} for the $y$-axis (chart \circled{2}); to rank countries by renewable percentage (chart \circled{4}), she partitions the data by year and uses \code{rank}; for charts \circled{5} and \circled{6}, she computes the global median using aggregation and merges the results with the previous table to surface all necessary fields.

\bpstart{Exploration with \df}
Using \df to complete the same analysis session, Megan's experience is quite different. Instead of transforming data and creating visualizations with code, Megan's main task is to describe visualization goals with UI interactions and NL inputs and ask the AI model to realize them.

\begin{figure*}[t]
    \centering
    \includegraphics[width=\linewidth]{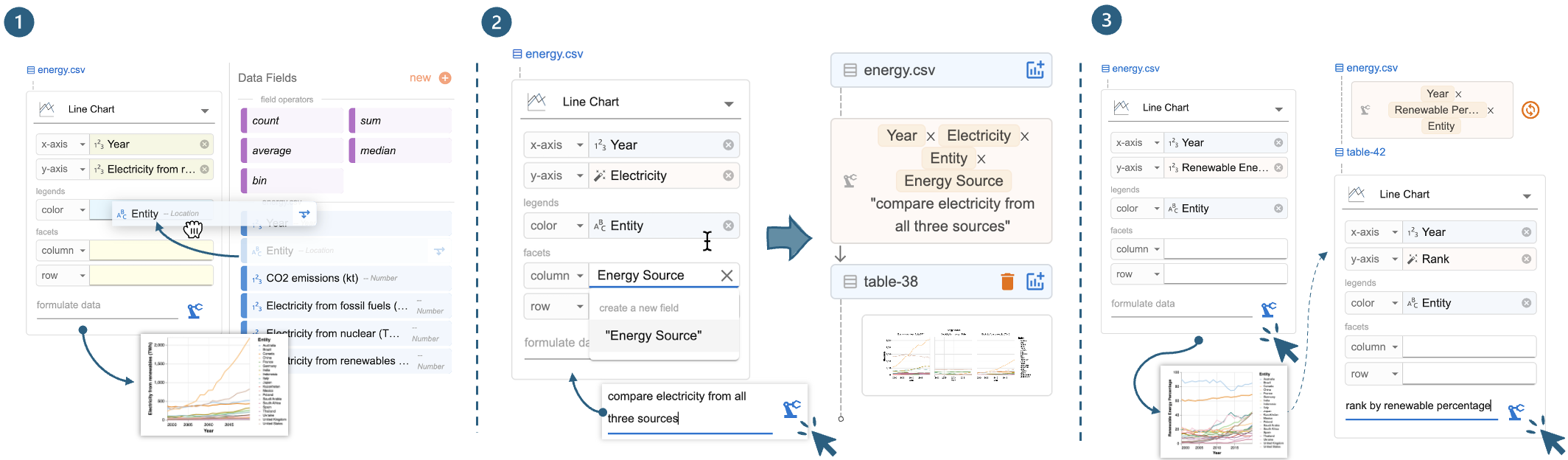}
    \caption{Experiences with \df: (1) creating the basic renewable energy chart using drag-and-drop to encode fields;  (2 and 3) creating charts requiring new fields by providing field names and optional natural language instructions to derive new data.}
    \label{fig:data-anvil-basics-facets}
\end{figure*}

Megan starts with basic line charts to visualize trends of electricity from renewables (\autoref{fig:example-analysis-session}-\circled{1}A). Since all three required fields are available from the input data, Megan simply selects the chart type ``line chart'' in the encoding shelf and drags and drops fields to their corresponding visual channels (\autoref{fig:data-anvil-basics-facets}-\circled{1}). \df then generates the desired visualization. To visualize the CO$_2$ emission trends, Megan swaps the $y$-axis encoding with \code{CO2 emissions (kt)}$\rightarrow y$.

Megan now needs to create the faceted line chart to compare electricity from all energy sources, which requires new fields \code{Electricity} and \code{Energy Source}. With \df, Megan can specify the chart using new data fields and NL instructions in the chart builder (\autoref{fig:data-anvil-overview}-2) and ask the AI to transform the data. 
As \autoref{fig:data-anvil-basics-facets}-\circled{2} shows, Megan first drags and drops existing fields \code{Year} and \code{Entity} to the $x$-axis and \textsf{color}, respectively. Then, she types in the names of new fields \code{Electricity} and \code{Energy Source} in the $y$-axis and \textsf{column}, respectively, to indicate to the AI agent that she expects two new fields to be derived for these properties. Finally, Megan provides an instruction, ``compare electricity from all three sources,'' to further clarify the intent and clicks the formulate button.
To create the chart, \df first generates a Vega-Lite spec skeleton from the encoding (to be completed based on information from the transformed data). It then summarizes the data, encodings, and NL instructions into a prompt to ask an AI model to generate data transformation code to prepare the data that fulfills all necessary fields, which is then used to instantiate the chart skeleton. After reviewing the generated chart and data, Megan is satisfied and moves to the next task. 
\df also updates data threads (\autoref{fig:data-anvil-overview}-\circled{5}) with the newly derived data and chart. With data threads, Megan can switch the iteration contexts to instruct the AI model to create a new chart either from scratch or reusing a previous result. 

\begin{figure*}[t]
    \centering
    \includegraphics[width=\linewidth]{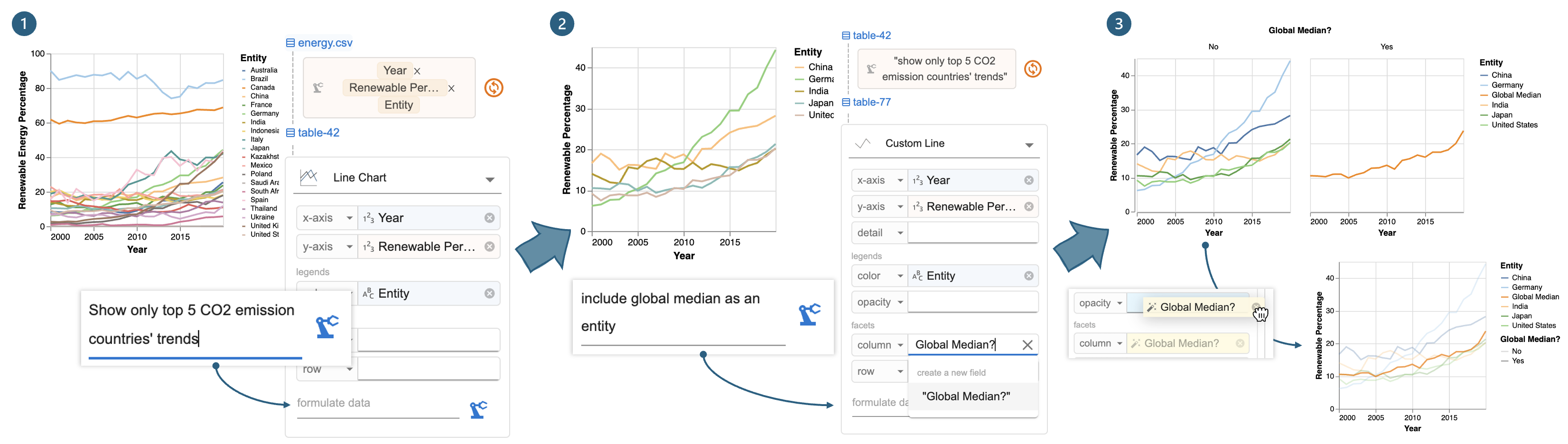}
    \caption{Iteration with \df: (1) provide an instruction to filter the renewable energy percentage chart by top CO$_2$ countries, (2) update the chart with \code{Global Median?} and instruct \df to add the global median alongside the top 5 CO$_2$ countries' trends, and (3) move \code{Global Median?} from \textsf{column} to \textsf{opacity} to update the chart design without deriving new data.}
    \label{fig:data-anvil-refinement}
\end{figure*}

Megan proceeds to visualize renewable energy percentage. Although it requires a different data transformation, Megan's experience is similar to the previous one: she drags-and-drops \code{Year} and \code{Entity} to $x$-axis and \textsf{color} (\autoref{fig:data-anvil-basics-facets}-\circled{3}), and enters the name of the new field ``\code{Renewable Energy Percentage}'' on the $y$-axis. Since Megan believes the field names are self-explanatory, she formulates the new data without an additional NL instruction. \df generates the desired visualization (\autoref{fig:data-anvil-refinement}-\circled{1}). To visualize the countries' renewable percentage ranks, building on the previous data, Megan adds a new field ``Rank'' to the $y$-axis and provides a short instruction. Because Megan builds the new chart on top of the previous data (note that in \autoref{fig:data-anvil-basics-facets}-\circled{3}, the chart builder box is positioned under the previous \textsf{table-42} as opposed to \textsf{energy.csv}), the AI model has more contextual information to correctly derive the renewable percentage rank (\autoref{fig:example-analysis-session}-\circled{4}) despite Megan's simple inputs.

Next, to visualize the renewable percentage trends of the top five CO$_2$ emitting countries, Megan decides to build on a previous chart to avoid creating a verbose prompt from scratch.
Megan first uses data threads (\autoref{fig:data-anvil-overview}-\circled{5}) to locate renewable percentage chart and opens it in the main panel. On top of that, Megan provides a new instruction below the local data thread, {``show only top 5 CO2 emission countries' trends,''} and clicks the ``derive'' button (\autoref{fig:data-anvil-refinement}-\circled{1}). \df updates the previous code to include a filter clause to produce the new data and visualization (\autoref{fig:data-anvil-refinement}-\circled{2}). Finally, to annotate the chart with global median trends, Megan forks a branch by copying the previous chart, as the new chart requires different encodings (and she wants to keep both visualizations available). Megan updates the visual encoding by (1) typing in a new field name \code{Global Median?} for \textsf{column} and (2) providing the edit instruction ``include global median as an entity'' (\autoref{fig:data-anvil-refinement}-\circled{2}). Once she clicks the derive button, \df generates the new chart (\autoref{fig:data-anvil-refinement}-\circled{3}). Upon inspection, Megan prefers to change the visualization type, with global average rendered in a different opacity as opposed to a different subplot. Since these two charts require the same data fields, Megan doesn't need to interact with the AI model --- she can directly update the design through the UI:  first selecting a new chart type ``custom line'' (which exposes more chart properties than the basic line chart) and moving \code{Global Median?} to the \textsf{opacity} channel. With all desired charts created, Megan concludes the analysis session. \autoref{fig:data-anvil-overview}-\circled{3} shows all the data threads from Megan.

\bpstart{Comparison of experiences}  These two tools offer different experiences and skill requirements for Megan to execute the analysis. However, both enable her to iteratively refine exploration goals and explore different branches to uncover insights.

The main difference between the two experiences is data transformation. In computation notebooks, Megan needs to prepare data for design updates, even seemingly small ones (e.g., charts-\circled{3} and \circled{5}). She must understand the data shape required and apply the correct transformations (e.g., unpivot for table~\circled{2}, join and union for table~\circled{6}). Proficiency in data transformation is essential for creating rich visualizations. In \df, Megan specifies high-level chart designs, and the AI implements the transformations. Regardless of the underlying data transformations, she conveys her intents uniformly through visual encodings (UI) and natural language inputs. Because Megan can use the shelf-configuration UI to specify chart design, the supplementary NL instruction is straightforward. Though Megan doesn't write code, \df provides artifacts like generated data, charts, and code with natural language explanations for her to review. By lowering the implementation skill barrier, \df allows users to focus more on analysis planning and reasoning.

Computation notebooks naturally support reuse. Megan can copy-edit previous code snippets or reuse variables to build new charts.  In \df, Megan directs the analysis using data threads. Megan can easily review the history and select previous results to instruct the AI model to create new charts from those contexts. This simplifies instructions to incremental updates, and the AI reuses previous outputs to avoid mistakes. If undesired results occur, she can backtrack and revise inputs using data threads (\autoref{fig:data-anvil-overview}-\circled{3}). Iteration isn't as easy with a chat-based tool. Iteration isn't as easy with a chat-based tool, where verbose prompts are needed to guide the AI and avoid unrelated histories.
\section{System Design}
\label{sec:system_design}
In this section, we present \df's system design.
First, to enable users to specify their intent using multiple paradigms (shelf-configuration UI and NL inputs) \df \textbf{decouples chart specification from data transformation}, solving them with template instantiation and AI code generation respectively.
Second, to support reuse, \df organizes \textbf{the iteration history as data threads with data as first-class objects}.
 This enables users to either locate a chart from a different branch and follow up or quickly revise and rerun the most recent instructions leading to the current chart. 
We will next detail how we implement these designs and explain how additional features help users understand AI-generated results.

\subsection{Composing charts from multi-modal inputs}
\begin{figure*}[t]
    \centering
    \includegraphics[width=1\linewidth]{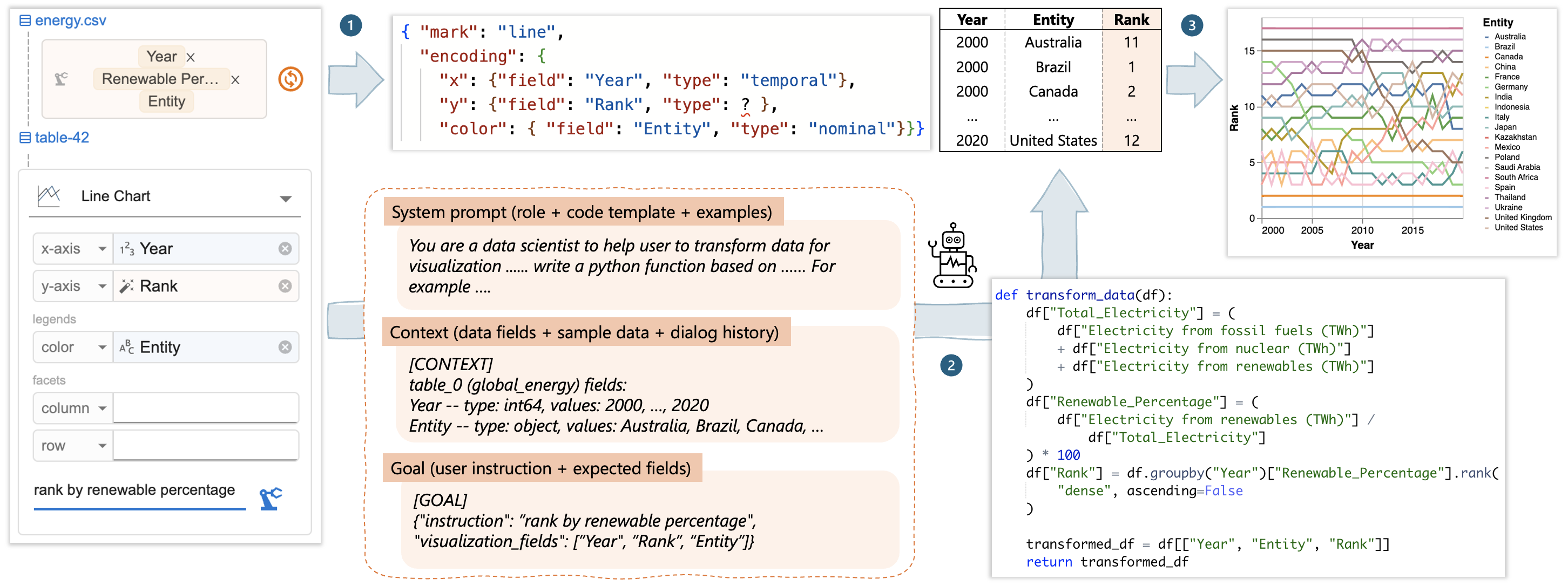}
    \caption{\df's workflow: (1) \df generates a Vega-Lite spec skeleton based on user specifications and chart type. (2) If new fields (e.g., \code{Rank}) are required, \df prompts its AI model to generate data transformation code. (3) The Vega-Lite skeleton is then instantiated with the new data to produce the desired chart.}
    \label{fig:multi-modal-ui-approach}
\end{figure*}

\autoref{fig:multi-modal-ui-approach} shows how \df decouples chart design and data transformation to support blended input methods. Given a user specification, \df generates the desired chart in three steps: (1) generate a Vega-Lite specification from the selected chart type, (2) compile a prompt and delegate data transformation to the AI, and (3) instantiate the Vega-Lite specification with the generated data.

\begin{figure*}
    \centering
    \includegraphics[width=1\linewidth]{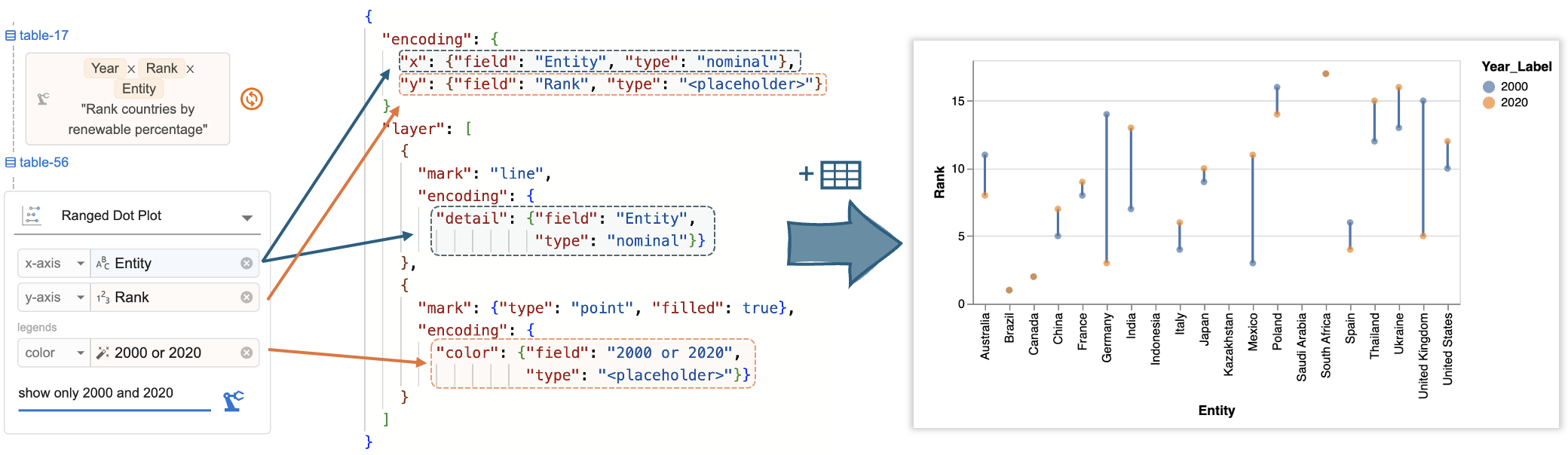}
    \caption{\df converts user encodings into a Vega-Lite specification, which is combined with AI-transformed data to visualize country ranks in 2000 and 2020.}
    \label{fig:template-instantiation}
\end{figure*}

\bpstart{Chart specification generation} \df adopts a chart type-based approach to represent visualizations, supporting five categories of charts: scatter (scatter plot, ranged dot plot), line (line chart, dotted line chart), bar (bar chart, stacked bar chart, grouped bar chart), statistics (histogram, heatmap, linear regression, boxplot) and custom (custom scatter, line, bar area, rectangle where all available visual channels are exposed). Each chart type is represented as a Vega-Lite template with a set of predefined visual channels, including position ($x$, $y$), legends (\textsf{color}, \textsf{size}, \textsf{shape}, \textsf{opacity}), and facet (\textsf{column}, \textsf{row}) that are shown to the user in the chart builder. For example, a line chart is represented as a Vega-Lite template \textsf{\{ "mark": "line", "encoding" : \{ "x": null, "y": null, "color": null, "column": null, "row": null\}\}}, and when the user selects line chart, channels $x$, $y$, \textsf{color}, \textsf{column}, and \textsf{row} are displayed in the chart builder. Chart type-based design enable \df to support predefined layered charts (e.g., ranged dot plot composed from line and scatter, \autoref{fig:template-instantiation}). Additional chart types (e.g., bullet chart) can be supported by adding Vega-Lite templates with respective channels to the library.

As the user inputs fields into the chart builder, either by dragging and dropping it from existing data fields or by typing in new fields they wish to visualize, \df instantiates the Vega-Lite template with provided fields. For example, as shown in \autoref{fig:multi-modal-ui-approach}-\filled{1}, when the user drags \code{Year}$\rightarrow x$, \code{Entity}$\rightarrow y$ and types \code{Rank} in $y$, the line chart template mentioned above is instantiated with provided fields: if the field is available in the current data table, both field name and encoding type are instantiated (e.g., \code{Year} with the temporal type), otherwise the encoding type is left as a ``<placeholder>'' to be instantiated later when data transformation completes. 
The shelf-configuration saves users efforts from writing prompts to explain complex chart designs. For example, to create a ranged dot plot--layered chart composed of scatter and line charts--the user only needs to fill the required fields in the UI. \df then populates corresponding fields in the predefined chart template (\autoref{fig:template-instantiation}). 

\bpstart{Data transformation with AI} From the chart builder, \df assembles a prompt and queries an LLM to generate python code to transform data. The data transformation prompt contains three segments: the system prompt, the data transformation context and the goal (illustrated \autoref{fig:multi-modal-ui-approach}-\filled{2}).

The \textbf{system prompt} describes the role of the LLM and the output format. Besides generic role descriptions (i.e., LLM as a data scientist for data transformation), the system prompt guides the LLM to solve the data transformation task in two steps. First, the LLM should refine the user's goal and output as a JSON object that elaborates intermediate and final fields to be computed from the original data. Then, the LLM should generate a python snippet following a provided template. The system prompt ends with an input-output example that illustrates the process. The design rationale behind the ``goal refinement'' step is to allow the LLM to reason about any potential discrepancy between users' provided fields and their instruction (e.g., users may ask about color by energy type but didn't put ``energy type'' on the color encoding) and determine the final list of fields to be computed.
\df then assembles \textbf{context prompts} that illustrate the data to be transformed, explaining the data fields by showing the data type and example values for each field, along with sample table rows. The data context provides valuable information related to data formats (such as data types, string formats, and whether columns contain null values) to the LLM, ensuring that the generated transformation code is executable on the given data. When a chart is specified based on previous results, the dialog history between \df and the LLM, including user instructions and previously generated code, is appended in context. This way, even if users' follow-up prompts is short, the grounded contexts help the model understand user intent and reuse previously generated code.
Finally, \df assembles a \textbf{goal prompt}, combining the NL instruction provided in the text box and field names used in the encodings. When users skip an NL instruction (\autoref{fig:data-anvil-basics-facets}-\filled{3}), the instruction part is left blank. This goal will be refined by the LLM (i.e., based on the system prompt) before attempting to generate the data transformation code.
With the full input, \df prompts the LLM to generate a response. Below shows the LLM's refined goal for the task in \autoref{fig:multi-modal-ui-approach}, and the generated code is shown in \autoref{fig:multi-modal-ui-approach}-\filled{2}.

\begin{imageonly}
\begin{center}
\begin{smpage}{0.95\linewidth}
\begin{lstlisting}[language=json]  
{ "detailed_instruction": "Calculate the percentage of electricity generated from renewables for each country per year. Then, rank the countries by their renewable percentage for each year.",
 "output_fields": ["Year", "Entity", "Renewable_Percentage", "Rank"],
 "visualization_fields": ["Year", "Rank", "Entity"],
 "reason": "To rank countries by their renewable percentage, we need to calculate the renewable percentage for each country per year and then determine the rank based on this percentage." }
\end{lstlisting}
\end{smpage}
\end{center}
\end{imageonly}

\smallskip

\df then runs the code on the input data. If the code executes without errors, the output data is used to instantiate the Vega-Lite script generated in the previous step. This is done by first inferring semantic types of newly generated columns (to determine their encoding type), and then assembling the data with the script to render the visualization (\autoref{fig:multi-modal-ui-approach}-\filled{3}). The generated code sometimes causes runtime errors due to an attempt to use libraries that are not imported, references to invalid columns names, or incorrect handling of \code{undefined} or \code{NaN} values. When such errors occur, \df tries to correct the errors by querying the LLM with the error message and a follow-up instruction to repair its mistakes~\cite{olausson2023self,chen2023teaching}. The visualization is generated when repair completes. \df updates the data threads upon creating the chart.

\subsection{Data threads}

\begin{figure*}
    \centering
    \includegraphics[width=1\linewidth]{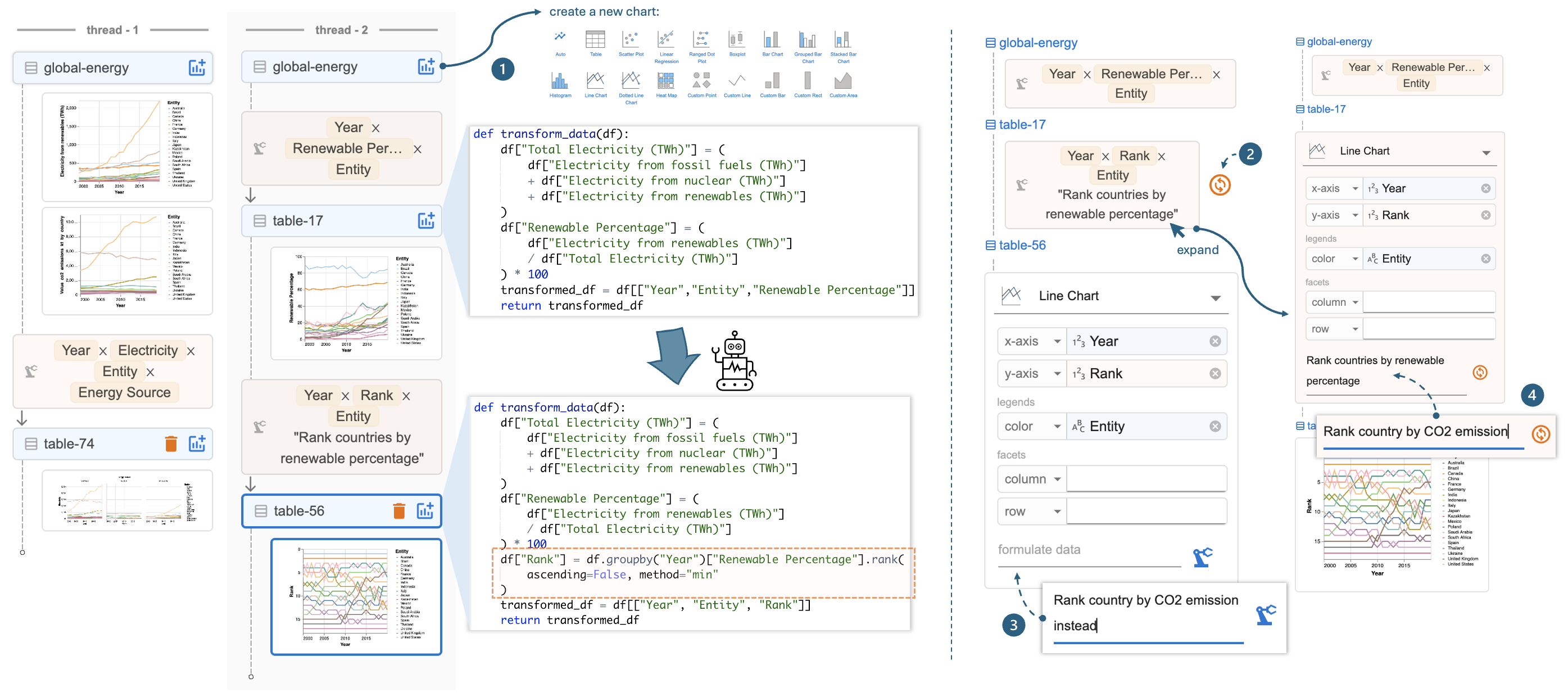}
    \caption{Data threads and local data threads (right). Users can select previous data or charts to create new branches, and the AI reuses code for new transformations based on user instructions. The local data thread offers shortcuts to (1) rerun the previous instruction, (2) issue a follow-up instruction, or (3) expand the previous card to revise and rerun the instruction.}
    \label{fig:data-threads-design}
\end{figure*}

Data threads visualize the analyst's interaction history with AI, allowing the analyst to control the iteration direction by selecting which data or chart the AI model should use to generate new charts.
In data threads, each node represents a version of the data, and these nodes are connected by edges that represent the user's instructions provided to the AI model for data transformation. Visualizations are attached to the data from which they were created. Centering the iteration history around data benefits user navigation because it reflects the sequence of user actions in creating these new data. 

When a user issues a follow-up instruction from an existing data or chart, \df provides the previous conversation history to the AI and instructs it to rewrite the code towards new goals. Each time the user forks a new branch using data threads, the authoring context switches automatically and is highlighted in the main panel for the user's awareness. This way, the AI model minimizes the risk of incorrectly using information from other branches for data transformation. As shown in \autoref{fig:data-threads-design}, the code and the conversation history are attached to each data node. In our design, when the user issues a follow-up instruction, the AI model generates new code by updating the previous code (which may involve additions, deletions, or both) to achieve the user's goal. This ensures that the code always takes the original data as the input, with all information accessible. This way, whether the user wants to update the data (e.g., {``now, calculate the average rank for each country''}), revise the previous computation (e.g., {``also consider nuclear as renewable energy''}) or create alternatives (e.g., {``rank by CO2 instead''}), the AI model can achieve these tasks as it has access to the full dialog history and the complete dataset. Note that an alternative design where we only pass current data to the AI model and ask it to write a new code to further transform it (i.e., reusing the data as opposed to reusing the computation leading to the data) would not be ideal. With  access to only the current data, this approach cannot handle ``backtracking'' or ``generating an alternative design'' styles of instructions effectively.

During iteration, analysts need to both (1) switch to different data or a chart far from the current one to explore a different direction and (2) perform quick follow-ups or revisions of the latest instruction based on the latest data. To accommodate these different needs, \df presents both global data threads and local data threads.
For global navigation, the key challenge is to help the user distinguish the desired content from others. To address this, data threads are located in a separate panel with previews of data, instructions, and charts to assist navigation (\autoref{fig:data-anvil-overview}). This supports users' differing navigation styles, whether they want to navigate by provenance (i.e., using instruction cards to locate desired data) or by artifacts (i.e., using visualization snapshots to recall data semantics). Once the user locates the desired data, they can click and open a previous chart, displaying it in the main panel. Additionally, they can create a new chart directly from the data~\autoref{fig:data-threads-design}-\filled{1}. In contrast, the local data thread is designed as part of the main authoring panel (\autoref{fig:data-anvil-overview}). It features a much-simplified view (i.e., hiding other visualizations created in this thread) to display a copy of the current thread in use. The main goals of the local data thread are to provide users with awareness of the current iteration context (so they don't need to cross-reference between the chart builder and the data threads panels) and to offer shortcuts for quick revisions of recently created charts.
As shown in \autoref{fig:data-threads-design}, the user can perform three types of revision tasks with local data threads: rerun the previous instruction (e.g., when the AI produces an incorrect result and they would like to quickly retry, \filled{2}), provide a follow-up instruction to refine the data (\filled{3}), and quickly open the previous instruction to modify and rerun the command (\filled{4}).

\begin{figure*}
    \centering
    \includegraphics[width=0.75\linewidth]{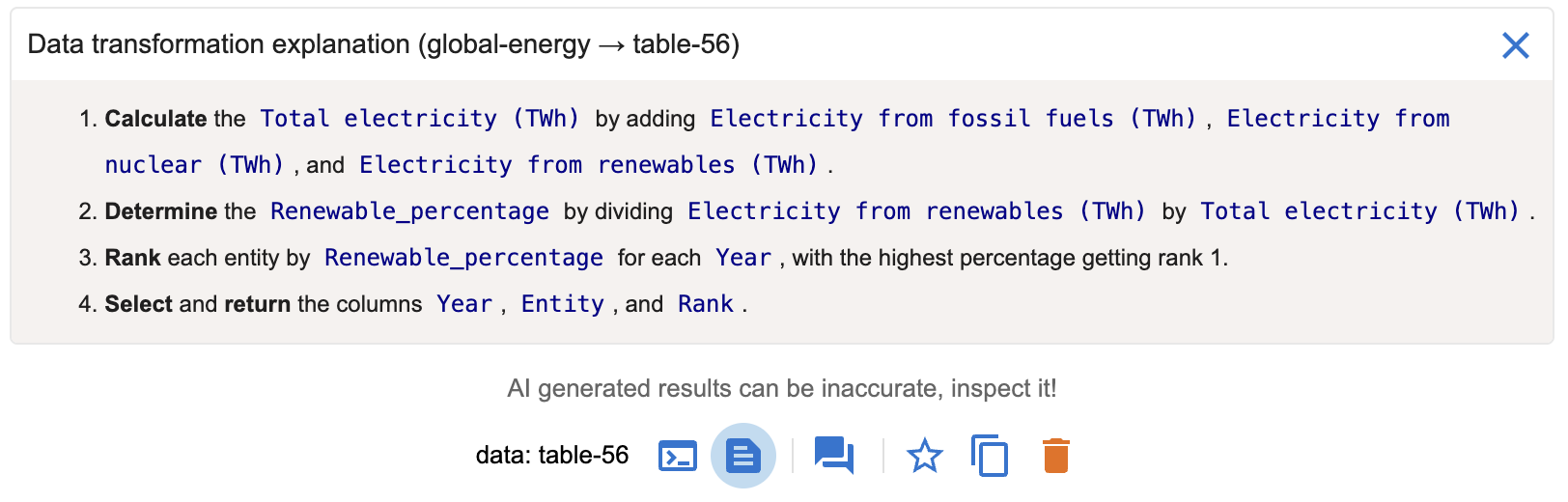}
    \caption{\df provides explanations of the code generated by AI to assist users understand the data transformation. This example is the explanation of the code behind table-56 in \autoref{fig:data-threads-design}.}
    \label{fig:code-explanation}
\end{figure*}

\subsection{Assisting user to inspect and style charts} 

As an AI-powered tool, \df allows users to verify AI-generated results and resolve AI's mistakes. It displays the transformed data and the visualization in the main panel and enables users to inspect generated code, its explanation, and the raw chat history through pop-up windows (\autoref{fig:data-anvil-overview}). This design accommodates various user verification styles~\cite{wang2021falx,gu2023analysts} such as viewing high-level correctness from the chart, inspecting corner cases in the data, examining the transformation output, and understanding the transformation process through the code. \df utilizes a code explanation module to help users understand the code, querying the AI model to translate code into step-by-step explanations. \autoref{fig:code-explanation} shows the explanation for the code behind table-56 in \autoref{fig:data-threads-design}. Expert users who would like to directly view the raw chat history between \df and the AI model (e.g., to inspect the LLM's raw reasoning process) can access this information from the ``view chat history'' pop-up window.
Note that despite that data transformations generated in the later iteration stages can be complex, users can verify its correctness against its predecessor because \df users create visualizations incrementally. This lowers users' verification efforts, as found in our study in \autoref{sec:evaluation}. To fix errors, users can take advantage of the data thread's iterative mechanism to rerun, follow up, or revise instructions. 

Benefiting from the decoupled chart specification and data transformation processes, when users want to update visualization styles (e.g., change color scheme, change sort order of an axis, or swap encodings) that do not require additional data transformation, they can directly perform edits in the chart builder. By updating channel properties or swapping encoded fields, these updates are directly reflected in the Vega-Lite script and rendered in the main panel. Unlike interactions with AI, which may have a slightly delayed response time, this approach allows users to achieve quick and precise edits with immediate visual feedback to refine the design.

\subsection{Implementation} \df is a React application with a python server for data transformation. \df has been tested with OpenAI models including GPT-3.5-turbo, GPT-4, GPT-4o, and GPT-4o-mini. We used GPT-3.5-turbo in our user study, and all but GPT-4 can generally response within 10 seconds. \df can sometimes be slow due to Vega-Lite rendering overhead (e.g., large datasets with more than 20,000 rows, long data threads with more than 20 charts). We envision that on-demand re-rendering of charts can improve its performance. 
\section{User Study Design}
\label{sec:evaluation}
To understand potential benefits and usability issues of \df, as well as users' interaction styles, we designed a user study that asks participants to reproduce exploratory data analysis sessions involving iteratively creating visualizations. 

\begin{figure}[t]
    \centering
    \includegraphics[width=\linewidth]{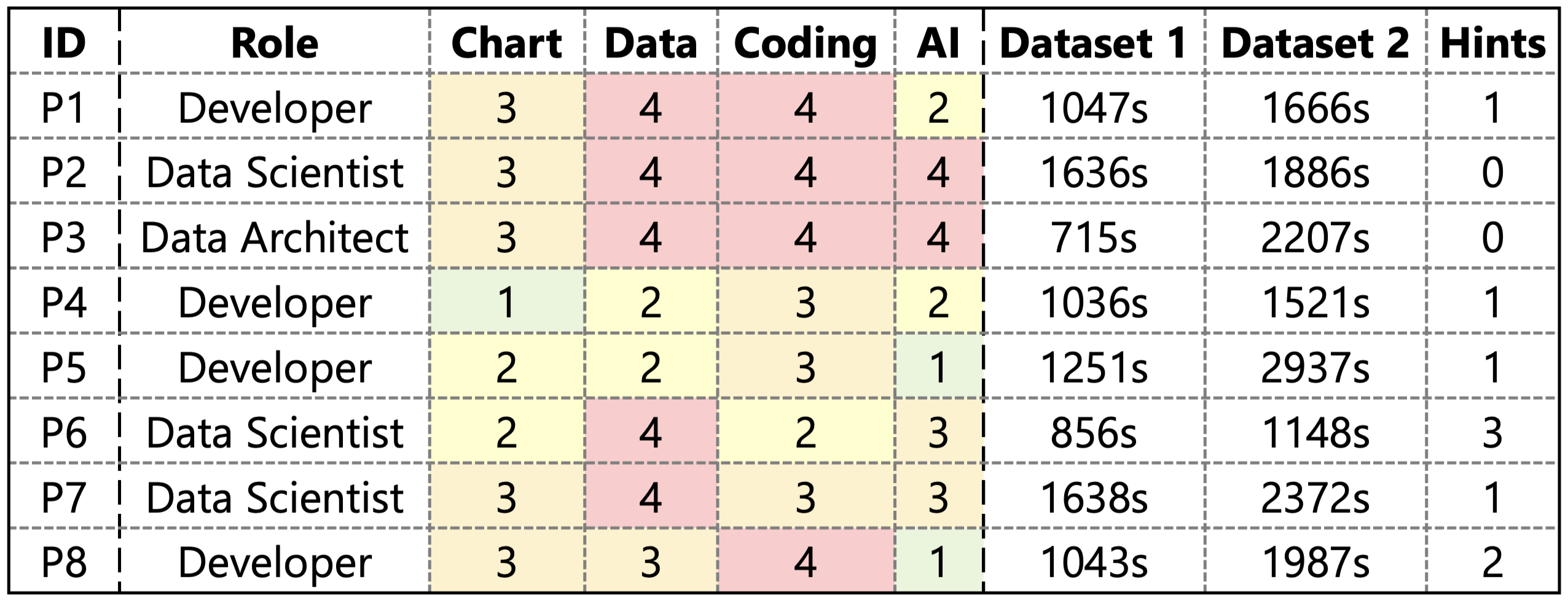}
    \caption{Participants' self-reported roles, expertise in chart creation, data transformation, programming, and AI assistants (1=novice, 4=expert), task completion time, and hints needed during study tasks.}
    \label{fig:participants}
\end{figure}

\bpstart{Participants} After piloting and refining the study design with three volunteers, we recruited eight participants from a large company. Participants self-rated their skills (\autoref{fig:participants}) on a scale of 1 to 4 (``Novice,'' ``Intermediate,'' ``Proficient,'' and ``Expert'') in: (1) chart creation -- experience with chart authoring tools or libraries, (2) data transformation -- experience with data transformation tools and library expertise, (3) programming, and (4) AI assistants -- experience with large language models (e.g., ChatGPT~\cite{achiam2023gpt}) and prompting. 

\bpstart{Setup and procedure} Each study session, conducted remotely with screen sharing, consisted of four sections within a 2-hour slot. After introduction, participants followed step-by-step instructions in the tutorial slides ($\sim$25 minutes). Participants then completed a practice task with the option to ask questions ($\sim$15 minutes) to test their understanding. Next, participants completed two study tasks, with only clarification questions allowed -- we recorded hints they requested. The two study tasks involved creating 16 visualizations, 12 requiring data transformation. Participants were encouraged to think aloud. We concluded with a debriefing to (1) compare participants' \df experiences with other tools, (2) understand their strategies using \df, and (3) gather impressions and suggestions for improvements. Breaks between phases were encouraged.

\bpstart{Tutorial and practice tasks} We used the global energy dataset (described in \autoref{sec:illustartive-scenarios}) for the tutorial and practice tasks. In the tutorial, participants followed detailed instructions to recreate the six visualizations from \autoref{fig:example-analysis-session} (all but chart \filled{4}).
In addition, participants also learned to inspect results and work with the AI's mistakes. 
In the practice tasks, participants were asked to do similar analyses but focusing on the electricity from nuclear power, they were further asked to create a bar chart to visualize the difference of energy produced from nuclear power between 2000 and 2020 for each country.

\begin{figure*}[t]
    \centering
    \includegraphics[width=1\linewidth]{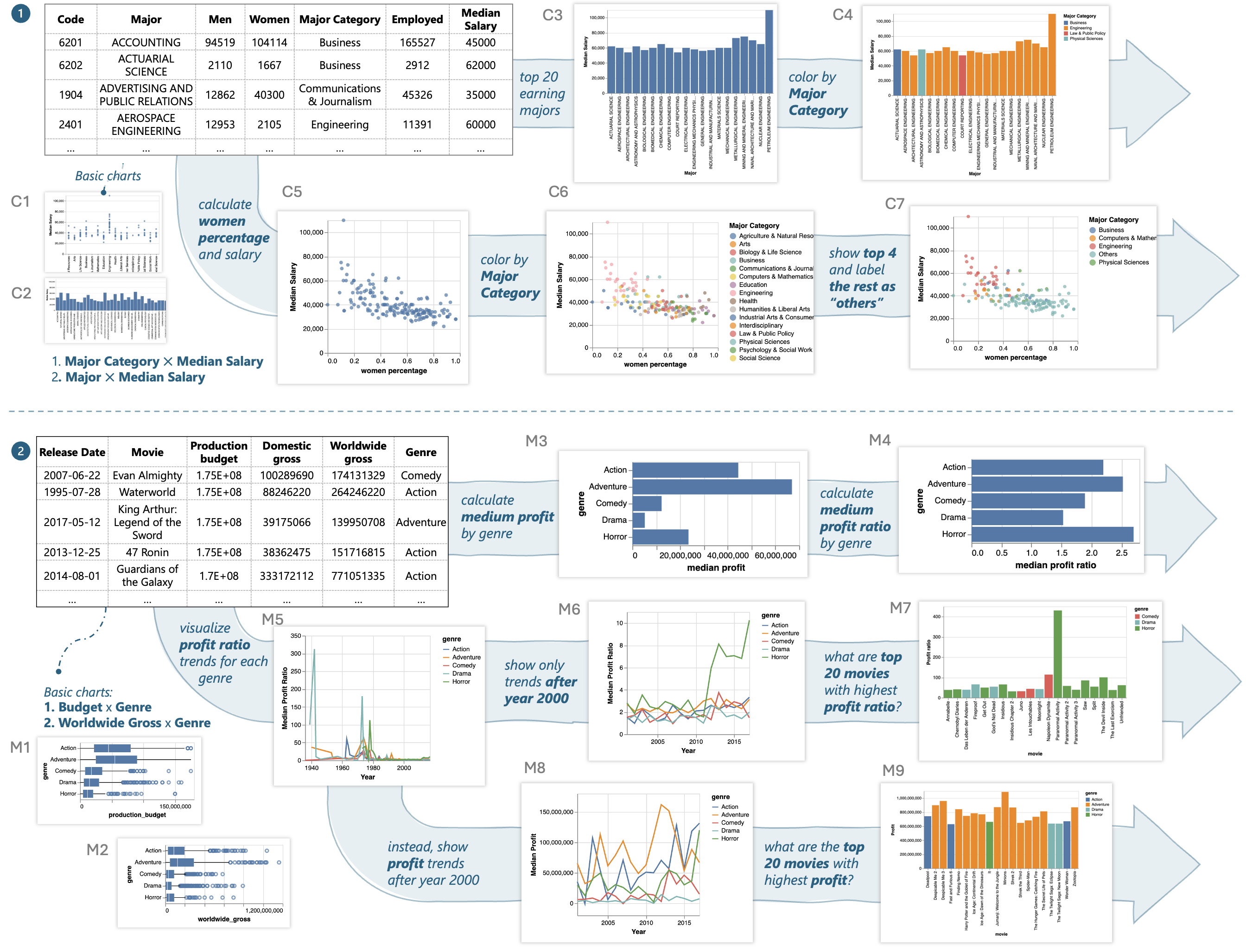}
    \caption{The dataset and tasks in our user study. (1) Dataset 1: Understanding top earning majors and the relation between salary and women percentage. (2) Dataset 2: Exploring movie genres with best return-on-investment values (profit vs. profit ratio) and top movies. The branching directions are added for illustration; participants developed their own iteration strategies. We refer to these target charts as C1-7 for the college dataset and M1-9 for the movies dataset.}
    \label{fig:study-tasks}
\end{figure*}
\bpstart{Study tasks} To focus on participants' iterative chart creation processes, rather than their ability to create a single chart or derive insights from exploration, we used an \emph{exploration session reproduction} approach. Participants were asked to reproduce two data exploration sessions conducted by an experienced data scientist. We wanted to see if participants could iteratively create charts with \df, without requiring them to come up with exploration objectives on the fly (otherwise we would limit our participants to highly skilled data scientists). We used two exploration sessions from David Robinson's live stream analysis of Tidy Tuesday datasets.

\autoref{fig:study-tasks}-\filled{1} shows the first data exploration session: given a dataset on college majors and income data (173 rows $\times$ 7 columns), participants were asked to create seven visualizations: two basic charts and five requiring data transformation. These visualizations progressively explored the top-earning majors and the relationship between gender ratio and major salary. 
The process required participants to derive new fields (e.g., gender ratio), filter data (e.g., top 20 earning majors), derive new data (e.g., derive top earning major categories), and perform conditional formatting (e.g., color by top 4 categories and "others"). We provided a task description and reference chart (like chart reproduction studies in \cite{ren2017chartaccent,ren2018reflecting}) for all but the last two visualizations. Without reference charts for the final two, we asked participants to verify correctness, probing their verification strategies. We did not provide iteration directions, letting participants develop iteration techniques.

\autoref{fig:study-tasks}-\filled{2} shows the second data exploration session: given a movie dataset with budget and gross information (3281 rows $\times$ 8 columns), participants created nine visualizations. These visualizations explored movies and genres with the highest return on investment, comparing profit and profit ratios. Besides two basic box plots showing budget and worldwide gross distribution, the other seven charts required data transformation, including calculation and aggregation (average profit and profit ratio for each genre), string processing (extract year for trends), filtering (year > 2000), and partitioning and ranking (top 20 movies for each metric). We hid references for the final two charts to probe participants' verification process. In the following, we use ``chart-C$k$'' and ``chart-M$k$'' to refer to the $k$-th target charts in \autoref{fig:study-tasks} for the college and movies datasets, respectively.





\definecolor{myblue}{HTML}{4DABF5}
\definecolor{myyellow}{HTML}{FFCD38}

\begin{figure*}
    \centering
    \includegraphics[width=\linewidth]{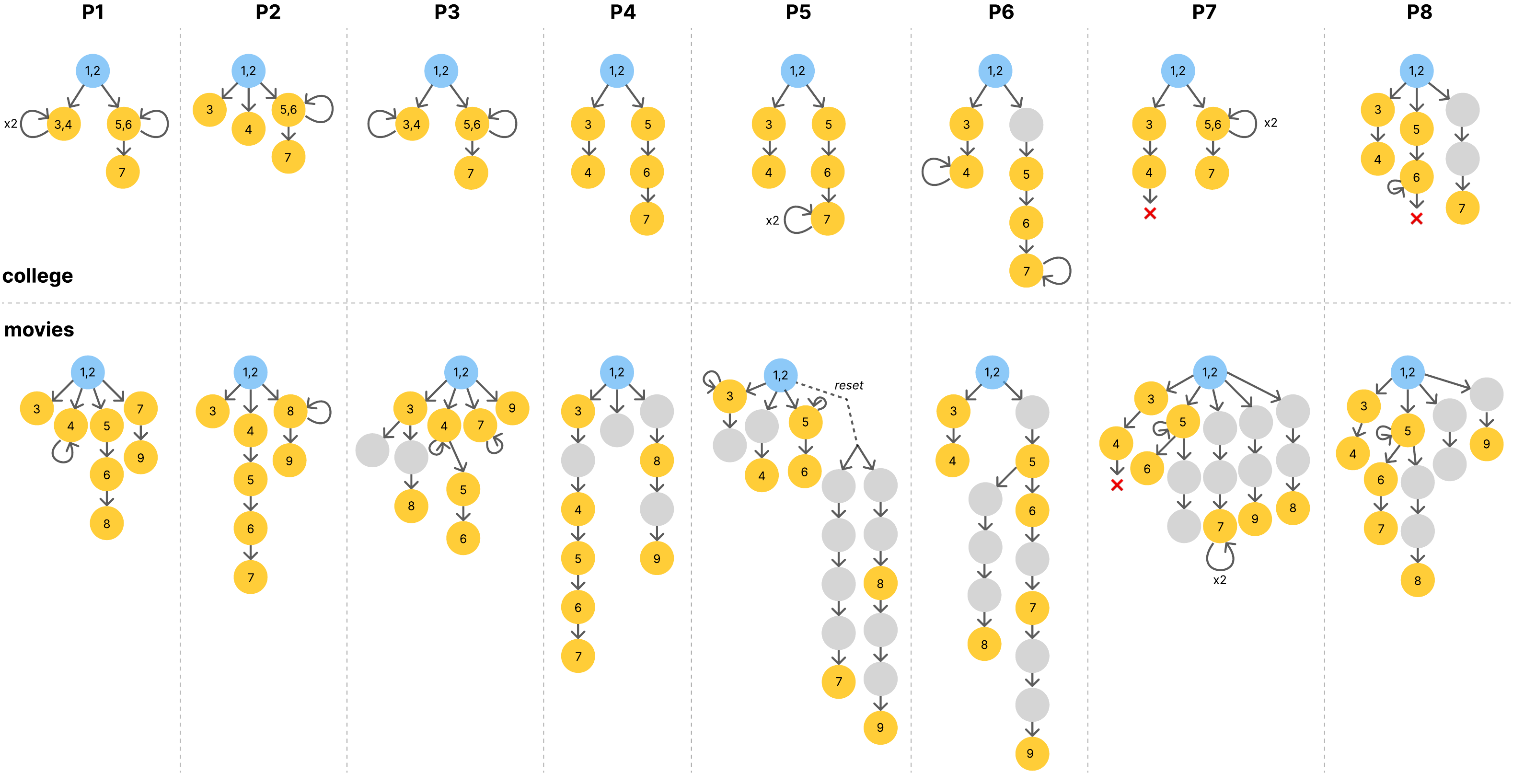}
    \caption{Participants' workflow for study tasks in \autoref{fig:study-tasks} (C1-7 for college, M1-9 for movie). Each node represents a data table version, with \hlc[myblue!64]{blue} for initial datasets, \hlc[myyellow]{yellow} for data tables instantiating (one or multiple) target visualizations in \autoref{fig:study-tasks} (number $i$ in the node indicate the $i$-th target visualizations for the given dataset), and \hlc[gray!30]{gray} for others. Self-loop arrows indicate prompt revisions and data table updates (`$\times$2' indicates two revisions).}
    \label{fig:user-workflows}
\end{figure*}

\section{User Study Results}

Here we report user study findings including users' task completion statistics as well as their prompting, iteration and verification styles. We highlight \pquote{user quotes} and \pprompt{example prompts} in this section.

\bpstart{Task completion} All participants successfully completed all 16 visualizations (\autoref{fig:participants}): participants took less than 20 mins on average to finish the seven charts in task 1, and about 33 mins for the nine charts in task 2. Since we let participants deviate from the main exploration task (e.g., in task 2, P4 asked to sort the bar chart for top profitable movies even though it was not required), the recorded completion time is an overestimate of the actual task time. During the study, six participants asked for hints to get unstuck during tasks; we categorize them as follows:
\begin{itemize}[leftmargin=*]
\item Task clarification: P1 didn't realize that top movies were restricted to movies after 2000; P4 and P6 required hints about the difference between profit and profit ratio in task 2; P6 asked about whether the $x$-axis should be \code{Year} or \code{Date} for movie profit trends.
\item Data clarification: P6 an P8 were prompted to notice the difference between fields \code{Major} and \code{Major Category} in task 1.
\item System performance: P5 encountered a performance issue when they created large sized charts. Tn task 2, P5 created multiple bar charts with \code{Movie} mapped to the $x$-axis, resulting in bar charts containing 1300 categorical values. They were advised to reset the exploration session and resume tasks.
\item Chart encoding: P7 and P8 required hints on ``why the chart didn't render color legends'' when they didn't put a field in the color encoding; they expected to specify it only in NL input but not in the concept encoding shelf.~\footnote{In the study version, \df didn't include the feature of resolving conflicts between the users' NL and encoding shelf inputs. This feature was introduced later.}
\end{itemize}

\noindent During the debriefing, participants commented that these tasks would be much more difficult to complete with tools they are familiar with. P1, a programming expert, mentioned that they were \pquote{``obviously much faster''} with \df as it helped with data transformations. When asked about their experience comparing against chat-based AI assistants, participants noted (1) the iteration support makes it easier to create more charts and (2) the UI + NL approach in \df is more effective for communicating intent structurally. For example, P2 mentioned \pquote{``with ChatGPT, I would have to put a bit more effort to specify the instructions to get what I want, iterations here is much faster with UI.''}
P4 mentioned that \pquote{``with ChatGPT, you need to give much more context, I need to describe in detail about what x,y-axes should be, but here I can just provide with UI,''} and further commented that UI + NL \pquote{``helped me in framing and structuring the different transformations that we need to do to get to that end result.''}

\bpstart{Iteration styles} Participants developed their own iteration styles working with \df--\autoref{fig:user-workflows} illustrates their organization of data threads in their workspaces upon completing the study tasks. Although our participant pool of 8 did not encompass all possible users' data exploration styles with \df, we observed surprising behavior clusters and distinct approach differences. We characterize participants' iteration styles based on their preferences between ``wider'' versus ``deeper'' tree structures, ``backtrack and revise'' versus ``follow up'' for providing new instructions to the AI, as well as their preferences for including intermediate tables in their threads.

\medskip

\noindent\emph{(Wide versus deep tree organizations):} From the high-level organization of data threads, one group of participants (P1, P3, P5, P7, P8) preferred to branch out more often with shorter data threads than the other group (P2, P4, P6), who preferred to create fewer but longer data threads instead.
P1 explained that their preference of more branches with shorter data threads came from their coding style of \pquote{``creating as many as transformation as I can from one single table without generating derived tables''} to keep the system's memory usage minimal and keep the workspace \pquote{``terse.''}
On the other hand, P2, who preferred longer data threads, mentioned \pquote{``I definitely like to be able to just work on top of that and like going forward by just giving a new prompt, because it remembers the context prior to the last one. It ends up generating the right data and visualization.''} P2 further commented that \pquote{``going back created too much branching''} and they preferred to use longer threads to just provide updates for \pquote{``smooth train of thoughts.''} To effectively work with long threads, P4 organized their exploration process thoughtfully, as they were \pquote{``using the prompts as my anchor, so, when I wanted to figure out where I wanted to go, it was the prompts that I was looking for.''} 

\medskip

\noindent\emph{(Backtracking versus following-up):} We observed interesting patterns in participants' preferences when creating new charts or correcting unexpected results: some preferred revising previous instructions (evident from workflows with more self-loop arrows), while others favored following up (characterized by more forward arrows and intermediate gray data nodes). The first group, represented by P1, P2 and P3, preferred to go back and re-issue prompts, either to enrich the previous data to support multiple target visualizations (indicated by yellow nodes with multiple target charts), or to update the data to correct unexpected results. For example, when P1 and P3 worked on coloring the top 20 earning majors with their major categories (chart-C4 in \autoref{fig:study-tasks}), they revised the previous prompt (\pprompt{``show only top 20 majors based on median salary''} $\rightarrow$ \pprompt{``show only top 20 majors based on median salary, include major category''} by P1) to include \code{Major\_Category} so that both old and the new charts can be created from the same data. To correct a mistake they made in creating chart-M7 (they forgot to instruct the AI to show only top 20 movies), P3 chose to go back and revise their previous prompt (\pprompt{``calculate the profit ratio per movie (worldwide\_gross/budget) after 2000''} $\rightarrow$
\pprompt{``calculate the profit ratio per movie (worldwide\_gross/budget) after 2000 and display the top 20 higher profit ratio movies''}). P1 commented that \pquote{``I don’t like to pollute my workspace''} and \pquote{``I like to keep my workspace as clean as possible.''} P3 mentioned that their preference of revision came from the concept of building a \pquote{``global expanded dataset''} so that \pquote{``[when I] need to calculate the new thing or see a new visual I can come back to the new expanded data set.''}

On the other hand, another group, represented by P4, P5, P6, and P7, preferred not only to issue follow-up instructions for new charts but also to provide updates with very brief instructions at each step, creating many intermediate nodes along the way (gray nodes in \autoref{fig:user-workflows}). For example, P5 created chart-M7 (top movies with highest profit ratio colored by genre) in five steps: \textit{``filter movies after year 2000''} $\rightarrow$ \textit{``show top 5 highest profit ratio''} $\rightarrow$ \textit{``bring back movie''} (i.e., the {Movie} field)  $\rightarrow$ \textit{``show top 10''} $\rightarrow$ \emph{``calculate profit ratio,''} creating four intermediate nodes. 
P5 noted that \pquote{``probably redoing would make sense, but if I can think that I can build on top of that, there is no value for me to go back and start from that, [which] kind of nullify these things [I have done],''} as they preferred to keep their work around. P6 mentioned that they adapt their iteration style based on the type of mistakes they encountered: \pquote{``if it is something intermediate where I've made the mistake, I'll go [create a new instruction] and fix the previous step''} but when it \pquote{``is a totally new kind of visualization I have in my mind''} or \pquote{``if it is something I missed altogether, I will just cancel the whole thing and start from scratch.''}

\medskip

\noindent\emph{(Choices of data to iterate on):} Participants had different strategies deciding which previous data/charts to use to create new charts. P1 chose to derive the new chart from a previous chart that shares similar visual design. For example, P1 created chart-M9 from M7 since they are both bar charts showing top ranked movies, despite one is based on profit while another is based on profit ratio. 
In a different fashion, P2, P4 and P5 often branch out based on similarity of computations used. For example, P2 created chart-M7 about movies with highest profit ratio based on chart-M6 showing profit ratio trends for each genre over time, as they shared the same computation ``profit ratio.'' 
P2 explained their data-centric approach was because they \pquote{``prefer to have more control over the data as opposed to the chart later on.''} They also appreciated that \df \pquote{``sort of brings together both data-centric and chart-centric people.''}

\bpstart{Prompt styles} Prompts created by participants are all short (less than 20 words). We observed that participants created diverse styles of prompts, both in terms of how they phrase the instruction (e.g., question, command) and the subject they asked (e.g., describing expected visual output or output data property, providing computation formula). The most common style of prompts is imperative commands, that either describe the transformation to be conducted or the property of the desired output. For example, to filter top earning movies, participants used prompts \pprompt{``show only top 20''} [P6] and\pprompt{``filter top 10 movies based on median profit''} [P5]. Participants also used command-style prompts for describing computations (e.g., \pprompt{``calculate ratio of worldwide\_gross by production\_budget''} [P5]) and for visual updates (e.g., \pprompt{``color by major category''} [P8]. 
We also observed that some participants prompted with questions (e.g., \pprompt{``can you show only the top 5 countries in terms of increases?''} [P7]), 
or prompted in a chat style (e.g., \pprompt{``Good. We need to now find the Median profit ratio each year for each genre''} [P2].

One participant, P5, had a distinct prompting style, that directly asked the AI to add, mutate, or retrieve columns on top of the previous data. For example, P5 asked \pprompt{``bring back major category''} to create chart-C4 from C3, \pprompt{``divide by 100,000''} for updating profit units, \pprompt{``bring back release\_date''} before they used a follow-up command \pprompt{``show only year greater than 2,000''} to filter movies by date. P7 preferred to use more verbose prompts to reiterate the computation they intended to achieve whenever they mentioned the concept. For example, to ensure that the AI would not interpret the computation differently, they copy/pasted the formula to the prompt whenever they mentioned profit ratio --- \pprompt{``median profit ratio (worldwide\_gross/production\_budget) by year and by genre.''} P6, on the other hand, preferred to use no additional prompts and provided more descriptive field names. For example, to create chart-C5, they mapped ``\code{percentage\_of\_women\_of\_Total\_Men\_and Women}'' to $x$-axis, \code{Median\_Salary} to $y$, and provided no prompt in the input box. In fact, we observed that \df can reliably transform data with self-explanatory field names (e.g., ``\code{renewable energy percentage}'', ``\code{women percentage}'', and ``\code{difference between 2020 and 2000}'') without any additional prompts. Some participants' preference for using shorter names and additional (short) prompts was \pquote{``to minimize the error space [for AI]''} [P7]. 

\bpstart{Verification} To proceed through iterative exploration, or repeat/correct a step, participants needed to understand the chart and verify that the transformation was performed correctly. Most of the time, participants spotted unintended output easily through incorrect patterns in rendered visualizations. This happened especially when there were differences in visual encoding (e.g., when P5 incorrectly mapped \code{release\_date} to the $x$-axis instead of \code{year} on chart-M5), cardinality (e.g., when P6 incorrectly asked the AI to color the bars by \code{major} instead of \code{major\_category} for chart-C6), or high-level patterns (e.g., when P7 requested \code{women} versus \code{median\_salary} for chart-C7, leading to results based on the count of women instead of the percentage). When the transformation is straightforward, participants visually inspected the chart and data to verify correctness. For example, after P3 asked \pprompt{``filter the year after 2000''} to show only profit ratio trends for movies after 2000 (chart-M8), they checked the $x$-axis domain and compared the generated chart with the pre-filtered one. Similarly, after P2 input \pprompt{``filter results to top 20 by major''} to find the highest earning majors (chart-C5), they referred to the previous chart with all of the majors' \code{median salary} sorted to check filtering correctness. 

To check whether unobvious computations were done correctly (e.g., whether the LLM computed profit ratio correctly), different participants' background impacted how they validated the results: participants either referred to (1) explanations of the code, (2) the actual code (even if they are non-python programmers), or (3) values in the result table to check correctness. P3 mentioned \pquote{``as an expert, I like to see the prompt to the model, and then the code generated; but as a business user, I would imagine using more data, chart, and explanations.''} while P4 commented \pquote{``[explanation] steps were really, really helpful in terms of figuring out whether it is doing the right thing as to what I'm asking it to do. That and also the data chart underneath.''} P7 noted that, for trust, the definition of a new field is more crucial than the actual code: \pquote{``I just want to make sure that definition, like profit ratio, when I check in, I only look at those definitions if they are correct. I'm less worried about the real coding piece.''} Thus, they use code explanations frequently to check definitions. Meanwhile, P7 stated that they felt some pressure from the study environment not to spend too much time understanding code for which they were not familiar with, but they would trust code more. We also observed participants who developed trust in a workflow (by examining code and data tables) when it was straightforward, and then, they assumed the more complicated transformations built on top of these steps worked.

\bpstart{Additional Feedback} Several users noted potential improvements of \df. P1 commented on how small interface variations might give different affordances. For instance, \pquote{``if there was a large view for data threads, it would encourage me to do more transformations and do more branching.''} P3 mentioned that they prefer the AI to ask the user to disambiguate when the intent is unclear rather than trying to solve the task with unclear specification. P7 used instructions that were very detailed and sometimes incorrect, which in turn, made iteration more difficult, since it was difficult to incrementally modify these instructions. We discussed the potential of having templates or AI feedback for instruction crafting to reduce errors.

\section{Discussion and Future Work}

\bpstart{Supporting recommendations in exploratory analysis} 
\df focuses on visualization authoring, where an AI completes tasks needed to achieve a user's intended action. We envision that \df can be enhanced with recommendation capabilities like Voyager~\cite{Wongsuphasawat2017voyager2}, Draco~\cite{moritz2018formalizing}, and Lux~\cite{lee2021lux} for suggesting visualization goals to help users ``cold start'' their analysis. \df's designs can benefit user experiences with visualization recommendation tools in two ways. First, because \df supports visualization beyond initial data formats, it overcomes the limitation of most existing tools, which only consider fields in the input table for recommendation. Second, \df's data threads provide a natural way for users to follow up the system's initial recommendations, either to dive deeper into an exploration direction, revising suggested charts, or to ask for different recommendations. To achieve this, we can add a recommendation component that can suggest a list of fields to be explored and let \df prepare data to surface the fields and create visualizations. The initial recommendation of the fields of interests can be generated either automatically from the analysis of input data characteristics or in a mixed-initiative approach, i.e., leveraging AI to generate them using a high-level natural language instruction provided by the user --- these fields do not have to be fields in input data, as \df can transform the data to derive them from existing ones. While \df's data transformation ability can extend the visualization exploration space, thus bringing in more potential insights to be discovered, it also increases the chances of suggesting field combinations that are either trivial, distracting, or even biased. Therefore, as part of future work, it would be valuable to explore ways to support visual recommendation in a larger exploration space, especially managing and communicating exploration paths to the user to prevent unintentional bias towards an undesired direction.

\bpstart{Coordinating data transformation and chart editing} \df derives new data based on users' inputs to instantiate the chart design, but it does not modify the chart itself (represented as a Vega-Lite specification). When the user wants to refine the chart design (e.g., updating color scheme or $x$-axis ordering), they edit it through GUI widgets in the encoding panel after the chart is created. This design leverages the natural and precise nature of UI updates, providing immediate visual feedback~\cite{vaithilingam2024dynavis}. It also utilizes current models' strengths in data transformation for more reliable outputs~\cite{gao2023text,lai2023ds}. In contrast, current models perform less effectively in editing charts or generating charts from data based on NL instructions, even when the data is prepared~\cite{chen2024viseval}. Despite this, some participants in the user study showed interest in asking \df to perform chart edits within the chart builder alongside data transformation. A potential solution is an agent-based system~\cite{wu2023autogen,zhang2024training} that plans whether to transform data, edit the chart script, or both based on user inputs, and dispatches agents to handle these tasks. The key challenge is managing response time and maintaining reliability, as AI agents often require multiple interactions to reach consensus. 


\bpstart{Asking users to clarify ambiguous inputs} \df adopts a generation verification approach: AI attempts to complete the user's request, and the user inspects the result to provide follow-up instructions. This interaction loop is enhanced by \df's local data thread design. There is an opportunity to make AI more proactive, such as actively seeking clarification from users when their inputs are ambiguous, before attempting to solve the task. This could reduce users' verification and revision efforts. For example, when the user issues an unclear request (e.g., \textit{``show top 5''}), the system can first analyze the goal, and then either present a refined goal for confirmation (e.g., \textit{``do you mean top 5 by renewable percentage?''}) or ask the users to clarify their intent (e.g., \textit{``what criteria should be used for ranking?''}). This proactive approach could also promote users' trust in the AI system. It is an interesting research direction to explore ways to prompt or train AI models to ask only necessary clarification questions, preventing users from being overwhelmed with low-level questions that might interrupt their workflow.

\bpstart{Study limitations} In our user study, we used the reproduction of professional data analysts' exploration sessions as the study tasks, rather than asking participants to perform free explorations. This choice was made to minimize the impact of participants' data analysis skills on their experience with \df, as our goal was not to assess their exploration skills. A follow-up study, where participants perform open exploration with their own data, can further investigate how \df can assist analysts with planning during exploration. In addition, as a limitation of our lab study, we could not capture users' longer-term learning effects. A future longitudinal study could further investigate how users' expectations with \df change over time and how this affects their specification styles and iteration strategies. 
\section{Related Work}
\label{related-work}

Compared to its predecessor, Data Formulator~\cite{wang2023data}, \df has transitioned from a single-turn chart authoring tool into an iterative visualization tool designed for data exploration. Concretely, Data Formulator~\cite{wang2023data} is a single-turn authoring tool that leverages different authoring paradigms for various types of data transformations. It uses programming-by-example for table reshaping and employs LLMs to generate code for single column derivation. However, users may struggle with choosing the appropriate paradigm for the required transformations. \df unifies the interaction paradigms with a blended UI and natural language input design, supporting iterative authoring. This allows users to build new charts from previous ones with minimal additional specification. \df's new interaction approaches not only broaden the expressiveness of supported data transformations but also reduce the users' specification overhead. We next illustrate related work on  chart authoring, data transformation, and data exploration tools that inspired the design of \df.

\bpstart{LLM-powered visualization tools} Large language models' code generation ability~\cite{achiam2023gpt,chen2021evaluating,lozhkov2024starcoder,touvron2023llama} motivates the designs of new AI-powered visualizations tools~\cite{dibia2019data2vis,maddigan2023chat2vis,tian2024chartgpt,wang2023data} that allows users to create visualization using high-level natural language descriptions. For example,  LIDA~\cite{dibia2023lida} can summarize data and use LLM to generate python code to generate visualizations. Because LLMs can struggle in understanding complex chart logic,  ChartGPT~\cite{tian2024chartgpt} decomposes visualization tasks into fine-grained reasoning pipelines (e.g., column selection, filtering, chart type selection, visual encoding), using chain-of-thoughts prompting~\cite{wei2022chain}. As single-turn interactive tools, they are not suitable for iterative analysis. For multi-turn interactions, users can directly chat with LLMs in Code Interpreter~\cite{achiam2023gpt} or Chat2Vis~\cite{maddigan2023chat2vis}. Code Interpreter equips the LLM with a Python interpreter so that the model can generate and execute code to transform data and create charts; Chat2Vis includes visualization-specific prompts to help the model generate visualizations more reliably. Since these tools organize the dialog linearly, users need to put in extra effort to clarify the context when there are branches, to reduce the chances of models applying incorrect contexts and making mistakes in the new task~\cite{liu2024lost,zhang2023tell,hsieh2024ruler}. 

\df is also an LLM-powered tool that shares similar prompt designs to LIDA and Chat2Vis (e.g., the use of data summaries) and supports NL interaction. The key difference is that \df blends UI and natural language inputs for chart specification, balancing precision and flexibility, rather than requiring users to describe everything in text. 
\df's data threads generalize linear contexts used in existing dialog systems, allowing users to control the iteration direction by providing authoring contexts to the AI model.

\bpstart{Other AI and synthesis-powered tools} Besides LLM-powered tools above, neural semantic parsing~\cite{mitra2022facilitating,narechania2020nl4dv,chen2022type}, and program synthesis-based tools~\cite{wang2021falx} have also been developed to address the visualization challenge. For example, NL4DV~\cite{narechania2020nl4dv} and NcNet~\cite{luo2021natural} leverage recurrent neural networks trained to translate NL queries into charts. NL2Vis~\cite{wu2022nl2viz} and Graphy~\cite{chen2022type} use a semantic parser to extract entities from the user's NL query and apply program synthesis algorithms to compose charts. Unlike LLMs, these tools are more restrictive in the supported data transformations and chart types, requiring very specific chart descriptions from the user.  While programming-by-examples (PBE) techniques are developed to tackle data reshaping challenges in chart authoring (e.g., Falx~\cite{wang2021falx} and Data Formulator~\cite{wang2023data}'s reshaping module), users need to prepare low-level examples to demonstrate the transformation intent, which deviates users from the high-level visualization workflow. 
For disambiguation, DataTone~\cite{DBLP:conf/uist/GaoDALK15} introduces disambiguation widgets for users to experiment with different entity extraction outputs for the generated query, and users can inspect paraphrased queries (in NL) to resolve ambiguity; Falx~\cite{wang2021falx} previews charts from multiple versions of data consistent with user examples. Benefiting from the use of LLMs, \df is more expressive. Inspired by how prior work displays candidate results and explains code to help users understand system outputs~\cite{DBLP:conf/uist/GaoDALK15,gu2023analysts,wang2023data}, \df displays generated code, data, chart and code explanation to assist user inspection. 

\bpstart{Visualization grammars and tools} The grammar of graphics~\cite{DBLP:books/daglib/0024564} inspired many modern visualization grammars (e.g., ggplot2~\cite{wickham2009ggplot2}, Vega-Lite~\cite{satyanarayan2017vegalite},
Altair~\cite{vanderplas2018altair}), where visualizations are mainly described by mappings from data columns to visual channels. Comparing to more expressive languages like D3~\cite{bostock2011d3} and Atlas~\cite{liu2021atlas}, high-level grammars hide the computation process of linking data items to visual objects to reduce visualization effort. Powered by these high-level grammars, interactive tools like Lyra~\cite{satyanarayan2014lyra}, Data Illustrator~\cite{liu2018data}, Charticulator~\cite{ren2019charticulator}, Tableau~\cite{stolte2002query}) have been introduced, where users leverage the shelf-configuration interface to specify visual encodings. To reduce authoring efforts, tools like Voyager~\cite{Wongsuphasawat2017voyager2}, Lux~\cite{lee2021lux}, and Draco~\cite{moritz2018formalizing} leverage rule and logic-based recommendation techniques to suggest visualizations from users' partial chart specifications. For example, Voyager lets users put a wildcard field into an encoding slot, and then automatically instantiates the wildcard field with different existing fields from the table, to produce interesting charts for users to explore. Note that these tools all require tidy input data~\cite{wickham2014tidy-data}, where all fields to be visualized should be data columns. Thus, users need to learn to use data transformation tools to prepare data~\cite{the_pandas_development_team_2023_7741580,wickham2019tidyverse,raman2001potter,kandel2011wrangler,polozov2015flashmeta,jin2017foofah,beth2020mage,huang2023interactive,chen2020multi}.

\df benefits from Vega-Lite's expressiveness to support rich visualization designs. \df inherits the shelf-configuration design from existing interactive tools and enhanced it with NL inputs for users to create charts that require data transformation.
While \df's custom fields resemble wildcard fields in Voyager~\cite{Wongsuphasawat2017voyager2}, they are semantically different: a custom field is for a field that users desire to visualize but not yet exist in the current table, requiring data transformation to surface, while a wildcard field refers to a field in the current table that the user does not specify explicitly. There is potential to unify these two as ``wildcard custom fields'' so that the system can recommend unspecified fields beyond the available fields in the current data (leveraging data transformation), which would broaden the exploration space.

\bpstart{Exploration history} Graphical history~\cite{heer2008graphical} and data provenance~\cite{buneman2001and} are essential in visualization authoring, especially in exploration tasks where branching and iterations are common. In computation notebooks, the exploration history is organized based on code blocks~\cite{mcnutt2023design,observable}. Data transformation tools like somnus~\cite{xiong2022visualizing} and Tableau Prep visualize data provenance based on transformation operators. Directed-graph models~\cite{shi2019task,kim2017graphscape} based on visual similarity are also used for visualization organization. To assist data scientists manage (messy) programming histories in computation notebooks, Verdant~\cite{kery2017variolite} introduces a design that visualizes users' edit histories of notebook and artifacts, allowing them to revisit different versions of the notebook; code gathering tools~\cite{head2019managing} leverage data dependency to extract a clean and minimal code snippet from a notebook that can reproduce a variable of interest. To support the management of different versions of code snippets created during the ideation process, Variolite~\cite{kery2017variolite} allows users to explicitly create branches when experimenting different implementations of a function and to switch among them later on. 

\df's data threads draw inspiration from these systems. The key difference is that data threads are designed for users to steer iteration directions by providing authoring contexts with AI. This approach organizes history around high-level user interactions with AI and hides operator-level details. We characterized users' interaction strategies based on their exploration tree~\cite{white2007investigating}. Provenance management techniques for notebooks can be applied to manage long data threads users created across different sessions (e.g., compressing long data threads into shorter ones with summaries). In the future, \df could render data threads as hierarchical trees~\cite{kim2017graphscape} to support navigation of large data threads in multiple granularities. Additionally, it could incorporate version toggles, similar to Variolite, allowing users to explore different versions of generated code more compactly, rather than presenting all exploration branches as separate data threads.


\bpstart{Multi-modal interaction} Despite natural language providing flexible and expressive interactions between human and AI, NL-only interaction is not always optimal for the users to clearly convey their intent, especially for conveying designs  pictured only in the user's  mind. To address this limitation, multi-modal models like ChatGPT~\cite{achiam2023gpt} and Gemini~\cite{reid2024gemini} have been introduced, allowing users to provide audios and images in their conversation with AI. New interactive tools are also developed to support multi-modal interaction. For example, DirectGPT~\cite{masson2024directgpt} allows users to directly point and click on a canvas to specify contexts or objects that NL instruction is based on to reduce prompting effort, Mage~\cite{beth2020mage} provides interactive widgets for users to control content in notebook, and DynaVis~\cite{vaithilingam2024dynavis} generates UI widgets dynamically based on user's NL inputs for chart editing with LLMs so that users can explore and repeat edits and see instant visual feedback from edits. \df's chart builder bridges the precision and affordance of GUI interaction with flexibility of NL inputs and thus exploits a multi-modal UI design for visualization authoring.

\section{Conclusion}
\label{sec:conclusion}

Visualization authors often create visualizations iteratively, alternating between data transformation and visualization steps. This process requires proficiency with tools and considerable effort to manage various versions of data and charts. 
Although AI-powered tools aim to reduce user effort, they fall short for iterative analysis, expecting users to specify their intent at once with NL inputs.
We present \df, an interactive system for iterative visualization authoring. \df features a multi-modal UI that allows users to specify visualizations using a blend of UI and NL inputs, enabling users to convey complex designs more precisely without verbose prompts.
To help user manage iteration directions, \df introduces data threads for users to navigate, branch, and reuse previous designs. In a user study with eight participants reproducing two challenging data exploration sessions consisting of 16 visualizations, we observed that \df enabled participants to develop their own iteration and verification strategies confidently with minimal hints.





\bibliographystyle{ACM-Reference-Format}
\bibliography{main}


\begin{thebibliography}{72}


\ifx \showCODEN    \undefined \def \showCODEN     #1{\unskip}     \fi
\ifx \showDOI      \undefined \def \showDOI       #1{#1}\fi
\ifx \showISBNx    \undefined \def \showISBNx     #1{\unskip}     \fi
\ifx \showISBNxiii \undefined \def \showISBNxiii  #1{\unskip}     \fi
\ifx \showISSN     \undefined \def \showISSN      #1{\unskip}     \fi
\ifx \showLCCN     \undefined \def \showLCCN      #1{\unskip}     \fi
\ifx \shownote     \undefined \def \shownote      #1{#1}          \fi
\ifx \showarticletitle \undefined \def \showarticletitle #1{#1}   \fi
\ifx \showURL      \undefined \def \showURL       {\relax}        \fi
\providecommand\bibfield[2]{#2}
\providecommand\bibinfo[2]{#2}
\providecommand\natexlab[1]{#1}
\providecommand\showeprint[2][]{arXiv:#2}

\bibitem[Achiam et~al\mbox{.}(2023)]%
        {achiam2023gpt}
\bibfield{author}{\bibinfo{person}{Josh Achiam}, \bibinfo{person}{Steven Adler}, \bibinfo{person}{Sandhini Agarwal}, \bibinfo{person}{Lama Ahmad}, \bibinfo{person}{Ilge Akkaya}, \bibinfo{person}{Florencia~Leoni Aleman}, \bibinfo{person}{Diogo Almeida}, \bibinfo{person}{Janko Altenschmidt}, \bibinfo{person}{Sam Altman}, \bibinfo{person}{Shyamal Anadkat}, {et~al\mbox{.}}} \bibinfo{year}{2023}\natexlab{}.
\newblock \showarticletitle{Gpt-4 technical report}.
\newblock \bibinfo{journal}{\emph{arXiv preprint arXiv:2303.08774}} (\bibinfo{year}{2023}).
\newblock


\bibitem[Barke et~al\mbox{.}(2023)]%
        {barke2023grounded}
\bibfield{author}{\bibinfo{person}{Shraddha Barke}, \bibinfo{person}{Michael~B James}, {and} \bibinfo{person}{Nadia Polikarpova}.} \bibinfo{year}{2023}\natexlab{}.
\newblock \showarticletitle{Grounded copilot: How programmers interact with code-generating models}.
\newblock \bibinfo{journal}{\emph{Proceedings of the ACM on Programming Languages}} \bibinfo{volume}{7}, \bibinfo{number}{OOPSLA1} (\bibinfo{year}{2023}), \bibinfo{pages}{85--111}.
\newblock


\bibitem[Bostock et~al\mbox{.}(2011)]%
        {bostock2011d3}
\bibfield{author}{\bibinfo{person}{Michael Bostock}, \bibinfo{person}{Vadim Ogievetsky}, {and} \bibinfo{person}{Jeffrey Heer}.} \bibinfo{year}{2011}\natexlab{}.
\newblock \showarticletitle{D{\({^3}\)} Data-Driven Documents}.
\newblock \bibinfo{journal}{\emph{{IEEE} Trans. Vis. Comput. Graph.}} \bibinfo{volume}{17}, \bibinfo{number}{12} (\bibinfo{year}{2011}), \bibinfo{pages}{2301--2309}.
\newblock
\urldef\tempurl%
\url{https://doi.org/10.1109/TVCG.2011.185}
\showDOI{\tempurl}


\bibitem[Buneman et~al\mbox{.}(2001)]%
        {buneman2001and}
\bibfield{author}{\bibinfo{person}{Peter Buneman}, \bibinfo{person}{Sanjeev Khanna}, {and} \bibinfo{person}{Tan Wang-Chiew}.} \bibinfo{year}{2001}\natexlab{}.
\newblock \showarticletitle{Why and where: A characterization of data provenance}. In \bibinfo{booktitle}{\emph{Database Theory—ICDT 2001: 8th International Conference London, UK, January 4--6, 2001 Proceedings 8}}. Springer, \bibinfo{pages}{316--330}.
\newblock


\bibitem[Chen et~al\mbox{.}(2021)]%
        {chen2021evaluating}
\bibfield{author}{\bibinfo{person}{Mark Chen}, \bibinfo{person}{Jerry Tworek}, \bibinfo{person}{Heewoo Jun}, \bibinfo{person}{Qiming Yuan}, \bibinfo{person}{Henrique~Pond{\'{e}} de Oliveira~Pinto}, \bibinfo{person}{Jared Kaplan}, \bibinfo{person}{Harrison Edwards}, \bibinfo{person}{Yuri Burda}, \bibinfo{person}{Nicholas Joseph}, \bibinfo{person}{Greg Brockman}, {et~al\mbox{.}}} \bibinfo{year}{2021}\natexlab{}.
\newblock \showarticletitle{Evaluating Large Language Models Trained on Code}.
\newblock \bibinfo{journal}{\emph{CoRR}}  \bibinfo{volume}{abs/2107.03374} (\bibinfo{year}{2021}).
\newblock
\showeprint[arXiv]{2107.03374}
\urldef\tempurl%
\url{https://arxiv.org/abs/2107.03374}
\showURL{%
\tempurl}


\bibitem[Chen et~al\mbox{.}(2024)]%
        {chen2024viseval}
\bibfield{author}{\bibinfo{person}{Nan Chen}, \bibinfo{person}{Yuge Zhang}, \bibinfo{person}{Jiahang Xu}, \bibinfo{person}{Kan Ren}, {and} \bibinfo{person}{Yuqing Yang}.} \bibinfo{year}{2024}\natexlab{}.
\newblock \showarticletitle{Viseval: A benchmark for data visualization in the era of large language models}.
\newblock \bibinfo{journal}{\emph{IEEE Transactions on Visualization and Computer Graphics}} (\bibinfo{year}{2024}).
\newblock


\bibitem[Chen et~al\mbox{.}(2022)]%
        {chen2022type}
\bibfield{author}{\bibinfo{person}{Qiaochu Chen}, \bibinfo{person}{Shankara Pailoor}, \bibinfo{person}{Celeste Barnaby}, \bibinfo{person}{Abby Criswell}, \bibinfo{person}{Chenglong Wang}, \bibinfo{person}{Greg Durrett}, {and} \bibinfo{person}{I{\c{s}}il Dillig}.} \bibinfo{year}{2022}\natexlab{}.
\newblock \showarticletitle{Type-directed synthesis of visualizations from natural language queries}.
\newblock \bibinfo{journal}{\emph{Proceedings of the ACM on Programming Languages}} \bibinfo{volume}{6}, \bibinfo{number}{OOPSLA2} (\bibinfo{year}{2022}), \bibinfo{pages}{532--559}.
\newblock


\bibitem[Chen et~al\mbox{.}(2020)]%
        {chen2020multi}
\bibfield{author}{\bibinfo{person}{Qiaochu Chen}, \bibinfo{person}{Xinyu Wang}, \bibinfo{person}{Xi Ye}, \bibinfo{person}{Greg Durrett}, {and} \bibinfo{person}{Isil Dillig}.} \bibinfo{year}{2020}\natexlab{}.
\newblock \showarticletitle{Multi-modal synthesis of regular expressions}. In \bibinfo{booktitle}{\emph{Proceedings of the 41st ACM SIGPLAN conference on programming language design and implementation}}. \bibinfo{pages}{487--502}.
\newblock


\bibitem[Chen et~al\mbox{.}(2023)]%
        {chen2023teaching}
\bibfield{author}{\bibinfo{person}{Xinyun Chen}, \bibinfo{person}{Maxwell Lin}, \bibinfo{person}{Nathanael Sch{\"a}rli}, {and} \bibinfo{person}{Denny Zhou}.} \bibinfo{year}{2023}\natexlab{}.
\newblock \showarticletitle{Teaching large language models to self-debug}.
\newblock \bibinfo{journal}{\emph{arXiv preprint arXiv:2304.05128}} (\bibinfo{year}{2023}).
\newblock


\bibitem[Dibia(2023)]%
        {dibia2023lida}
\bibfield{author}{\bibinfo{person}{Victor Dibia}.} \bibinfo{year}{2023}\natexlab{}.
\newblock \showarticletitle{LIDA: A Tool for Automatic Generation of Grammar-Agnostic Visualizations and Infographics using Large Language Models}.
\newblock \bibinfo{journal}{\emph{arXiv preprint arXiv:2303.02927}} (\bibinfo{year}{2023}).
\newblock


\bibitem[Dibia and Demiralp(2019)]%
        {dibia2019data2vis}
\bibfield{author}{\bibinfo{person}{Victor Dibia} {and} \bibinfo{person}{{\c{C}}a{\u{g}}atay Demiralp}.} \bibinfo{year}{2019}\natexlab{}.
\newblock \showarticletitle{Data2vis: Automatic generation of data visualizations using sequence-to-sequence recurrent neural networks}.
\newblock \bibinfo{journal}{\emph{IEEE computer graphics and applications}} \bibinfo{volume}{39}, \bibinfo{number}{5} (\bibinfo{year}{2019}), \bibinfo{pages}{33--46}.
\newblock


\bibitem[Gao et~al\mbox{.}(2023)]%
        {gao2023text}
\bibfield{author}{\bibinfo{person}{Dawei Gao}, \bibinfo{person}{Haibin Wang}, \bibinfo{person}{Yaliang Li}, \bibinfo{person}{Xiuyu Sun}, \bibinfo{person}{Yichen Qian}, \bibinfo{person}{Bolin Ding}, {and} \bibinfo{person}{Jingren Zhou}.} \bibinfo{year}{2023}\natexlab{}.
\newblock \showarticletitle{Text-to-sql empowered by large language models: A benchmark evaluation}.
\newblock \bibinfo{journal}{\emph{arXiv preprint arXiv:2308.15363}} (\bibinfo{year}{2023}).
\newblock


\bibitem[Gao et~al\mbox{.}(2015)]%
        {DBLP:conf/uist/GaoDALK15}
\bibfield{author}{\bibinfo{person}{Tong Gao}, \bibinfo{person}{Mira Dontcheva}, \bibinfo{person}{Eytan Adar}, \bibinfo{person}{Zhicheng Liu}, {and} \bibinfo{person}{Karrie~G. Karahalios}.} \bibinfo{year}{2015}\natexlab{}.
\newblock \showarticletitle{DataTone: Managing Ambiguity in Natural Language Interfaces for Data Visualization}. In \bibinfo{booktitle}{\emph{Proceedings of the 28th Annual {ACM} Symposium on User Interface Software {\&} Technology, {UIST} 2015, Charlotte, NC, USA, November 8-11, 2015}}, \bibfield{editor}{\bibinfo{person}{Celine Latulipe}, \bibinfo{person}{Bjoern Hartmann}, {and} \bibinfo{person}{Tovi Grossman}} (Eds.). \bibinfo{publisher}{{ACM}}, \bibinfo{pages}{489--500}.
\newblock
\urldef\tempurl%
\url{https://doi.org/10.1145/2807442.2807478}
\showDOI{\tempurl}


\bibitem[Gu et~al\mbox{.}(2023)]%
        {gu2023analysts}
\bibfield{author}{\bibinfo{person}{Ken Gu}, \bibinfo{person}{Ruoxi Shang}, \bibinfo{person}{Tim Althoff}, \bibinfo{person}{Chenglong Wang}, {and} \bibinfo{person}{Steven~M Drucker}.} \bibinfo{year}{2023}\natexlab{}.
\newblock \showarticletitle{How Do Analysts Understand and Verify AI-Assisted Data Analyses?}
\newblock \bibinfo{journal}{\emph{arXiv preprint arXiv:2309.10947}} (\bibinfo{year}{2023}).
\newblock


\bibitem[Head et~al\mbox{.}(2019)]%
        {head2019managing}
\bibfield{author}{\bibinfo{person}{Andrew Head}, \bibinfo{person}{Fred Hohman}, \bibinfo{person}{Titus Barik}, \bibinfo{person}{Steven~M Drucker}, {and} \bibinfo{person}{Robert DeLine}.} \bibinfo{year}{2019}\natexlab{}.
\newblock \showarticletitle{Managing messes in computational notebooks}. In \bibinfo{booktitle}{\emph{Proceedings of the 2019 CHI Conference on Human Factors in Computing Systems}}. \bibinfo{pages}{1--12}.
\newblock


\bibitem[Heer et~al\mbox{.}(2008)]%
        {heer2008graphical}
\bibfield{author}{\bibinfo{person}{Jeffrey Heer}, \bibinfo{person}{Jock Mackinlay}, \bibinfo{person}{Chris Stolte}, {and} \bibinfo{person}{Maneesh Agrawala}.} \bibinfo{year}{2008}\natexlab{}.
\newblock \showarticletitle{Graphical histories for visualization: Supporting analysis, communication, and evaluation}.
\newblock \bibinfo{journal}{\emph{IEEE transactions on visualization and computer graphics}} \bibinfo{volume}{14}, \bibinfo{number}{6} (\bibinfo{year}{2008}), \bibinfo{pages}{1189--1196}.
\newblock


\bibitem[Hsieh et~al\mbox{.}(2024)]%
        {hsieh2024ruler}
\bibfield{author}{\bibinfo{person}{Cheng-Ping Hsieh}, \bibinfo{person}{Simeng Sun}, \bibinfo{person}{Samuel Kriman}, \bibinfo{person}{Shantanu Acharya}, \bibinfo{person}{Dima Rekesh}, \bibinfo{person}{Fei Jia}, {and} \bibinfo{person}{Boris Ginsburg}.} \bibinfo{year}{2024}\natexlab{}.
\newblock \showarticletitle{RULER: What's the Real Context Size of Your Long-Context Language Models?}
\newblock \bibinfo{journal}{\emph{arXiv preprint arXiv:2404.06654}} (\bibinfo{year}{2024}).
\newblock


\bibitem[Huang et~al\mbox{.}(2023)]%
        {huang2023interactive}
\bibfield{author}{\bibinfo{person}{Yanwei Huang}, \bibinfo{person}{Yunfan Zhou}, \bibinfo{person}{Ran Chen}, \bibinfo{person}{Changhao Pan}, \bibinfo{person}{Xinhuan Shu}, \bibinfo{person}{Di Weng}, {and} \bibinfo{person}{Yingcai Wu}.} \bibinfo{year}{2023}\natexlab{}.
\newblock \showarticletitle{Interactive table synthesis with natural language}.
\newblock \bibinfo{journal}{\emph{IEEE Transactions on Visualization and Computer Graphics}} (\bibinfo{year}{2023}).
\newblock


\bibitem[Jin et~al\mbox{.}(2017)]%
        {jin2017foofah}
\bibfield{author}{\bibinfo{person}{Zhongjun Jin}, \bibinfo{person}{Michael~R. Anderson}, \bibinfo{person}{Michael~J. Cafarella}, {and} \bibinfo{person}{H.~V. Jagadish}.} \bibinfo{year}{2017}\natexlab{}.
\newblock \showarticletitle{Foofah: Transforming Data By Example}. In \bibinfo{booktitle}{\emph{{SIGMOD} Conference 2017, Chicago, IL, USA, May 14-19, 2017}}, \bibfield{editor}{\bibinfo{person}{Semih Salihoglu}, \bibinfo{person}{Wenchao Zhou}, \bibinfo{person}{Rada Chirkova}, \bibinfo{person}{Jun Yang}, {and} \bibinfo{person}{Dan Suciu}} (Eds.). \bibinfo{publisher}{{ACM}}, \bibinfo{pages}{683--698}.
\newblock
\urldef\tempurl%
\url{https://doi.org/10.1145/3035918.3064034}
\showDOI{\tempurl}


\bibitem[Kandel et~al\mbox{.}(2011)]%
        {kandel2011wrangler}
\bibfield{author}{\bibinfo{person}{Sean Kandel}, \bibinfo{person}{Andreas Paepcke}, \bibinfo{person}{Joseph Hellerstein}, {and} \bibinfo{person}{Jeffrey Heer}.} \bibinfo{year}{2011}\natexlab{}.
\newblock \showarticletitle{Wrangler: Interactive visual specification of data transformation scripts}. In \bibinfo{booktitle}{\emph{Proceedings of the ACM Conference on Human Factors in Computing Systems (CHI)}}. \bibinfo{pages}{3363--3372}.
\newblock
\urldef\tempurl%
\url{https://doi.org/10.1145/1978942.1979444}
\showDOI{\tempurl}


\bibitem[Kery et~al\mbox{.}(2017)]%
        {kery2017variolite}
\bibfield{author}{\bibinfo{person}{Mary~Beth Kery}, \bibinfo{person}{Amber Horvath}, {and} \bibinfo{person}{Brad~A Myers}.} \bibinfo{year}{2017}\natexlab{}.
\newblock \showarticletitle{Variolite: Supporting Exploratory Programming by Data Scientists.}. In \bibinfo{booktitle}{\emph{CHI}}, Vol.~\bibinfo{volume}{10}. \bibinfo{pages}{3025453--3025626}.
\newblock


\bibitem[Kery et~al\mbox{.}(2020)]%
        {beth2020mage}
\bibfield{author}{\bibinfo{person}{Mary~Beth Kery}, \bibinfo{person}{Donghao Ren}, \bibinfo{person}{Fred Hohman}, \bibinfo{person}{Dominik Moritz}, \bibinfo{person}{Kanit Wongsuphasawat}, {and} \bibinfo{person}{Kayur Patel}.} \bibinfo{year}{2020}\natexlab{}.
\newblock \showarticletitle{Mage: Fluid Moves Between Code and Graphical Work in Computational Notebooks}. In \bibinfo{booktitle}{\emph{Proceedings of the 33rd Annual ACM Symposium on User Interface Software and Technology}} (Virtual Event, USA) \emph{(\bibinfo{series}{UIST '20})}. \bibinfo{publisher}{Association for Computing Machinery}, \bibinfo{address}{New York, NY, USA}, \bibinfo{pages}{140–151}.
\newblock
\showISBNx{9781450375146}
\urldef\tempurl%
\url{https://doi.org/10.1145/3379337.3415842}
\showDOI{\tempurl}


\bibitem[Kim et~al\mbox{.}(2017)]%
        {kim2017graphscape}
\bibfield{author}{\bibinfo{person}{Younghoon Kim}, \bibinfo{person}{Kanit Wongsuphasawat}, \bibinfo{person}{Jessica Hullman}, {and} \bibinfo{person}{Jeffrey Heer}.} \bibinfo{year}{2017}\natexlab{}.
\newblock \showarticletitle{Graphscape: A model for automated reasoning about visualization similarity and sequencing}. In \bibinfo{booktitle}{\emph{Proceedings of the 2017 CHI conference on human factors in computing systems}}. \bibinfo{pages}{2628--2638}.
\newblock


\bibitem[Lai et~al\mbox{.}(2023)]%
        {lai2023ds}
\bibfield{author}{\bibinfo{person}{Yuhang Lai}, \bibinfo{person}{Chengxi Li}, \bibinfo{person}{Yiming Wang}, \bibinfo{person}{Tianyi Zhang}, \bibinfo{person}{Ruiqi Zhong}, \bibinfo{person}{Luke Zettlemoyer}, \bibinfo{person}{Wen-tau Yih}, \bibinfo{person}{Daniel Fried}, \bibinfo{person}{Sida Wang}, {and} \bibinfo{person}{Tao Yu}.} \bibinfo{year}{2023}\natexlab{}.
\newblock \showarticletitle{DS-1000: A natural and reliable benchmark for data science code generation}. In \bibinfo{booktitle}{\emph{International Conference on Machine Learning}}. PMLR, \bibinfo{pages}{18319--18345}.
\newblock


\bibitem[Lee et~al\mbox{.}(2021)]%
        {lee2021lux}
\bibfield{author}{\bibinfo{person}{Doris Jung~Lin Lee}, \bibinfo{person}{Dixin Tang}, \bibinfo{person}{Kunal Agarwal}, \bibinfo{person}{Thyne Boonmark}, \bibinfo{person}{Caitlyn Chen}, \bibinfo{person}{Jake Kang}, \bibinfo{person}{Ujjaini Mukhopadhyay}, \bibinfo{person}{Jerry Song}, \bibinfo{person}{Micah Yong}, \bibinfo{person}{Marti~A. Hearst}, {and} \bibinfo{person}{Aditya~G. Parameswaran}.} \bibinfo{year}{2021}\natexlab{}.
\newblock \showarticletitle{Lux: Always-on Visualization Recommendations for Exploratory Dataframe Workflows}.
\newblock \bibinfo{journal}{\emph{Proc. {VLDB} Endow.}} \bibinfo{volume}{15}, \bibinfo{number}{3} (\bibinfo{year}{2021}), \bibinfo{pages}{727--738}.
\newblock
\urldef\tempurl%
\url{https://doi.org/10.14778/3494124.3494151}
\showDOI{\tempurl}


\bibitem[Liu et~al\mbox{.}(2024)]%
        {liu2024lost}
\bibfield{author}{\bibinfo{person}{Nelson~F Liu}, \bibinfo{person}{Kevin Lin}, \bibinfo{person}{John Hewitt}, \bibinfo{person}{Ashwin Paranjape}, \bibinfo{person}{Michele Bevilacqua}, \bibinfo{person}{Fabio Petroni}, {and} \bibinfo{person}{Percy Liang}.} \bibinfo{year}{2024}\natexlab{}.
\newblock \showarticletitle{Lost in the middle: How language models use long contexts}.
\newblock \bibinfo{journal}{\emph{Transactions of the Association for Computational Linguistics}}  \bibinfo{volume}{12} (\bibinfo{year}{2024}), \bibinfo{pages}{157--173}.
\newblock


\bibitem[Liu et~al\mbox{.}(2021)]%
        {liu2021atlas}
\bibfield{author}{\bibinfo{person}{Zhicheng Liu}, \bibinfo{person}{Chen Chen}, \bibinfo{person}{Francisco Morales}, {and} \bibinfo{person}{Yishan Zhao}.} \bibinfo{year}{2021}\natexlab{}.
\newblock \showarticletitle{Atlas: Grammar-based Procedural Generation of Data Visualizations}. In \bibinfo{booktitle}{\emph{2021 {IEEE} Visualization Conference, 2021 - Short Papers, New Orleans, LA, USA, October 24-29, 2021}}. \bibinfo{publisher}{{IEEE}}, \bibinfo{pages}{171--175}.
\newblock
\urldef\tempurl%
\url{https://doi.org/10.1109/VIS49827.2021.9623315}
\showDOI{\tempurl}


\bibitem[Liu et~al\mbox{.}(2018)]%
        {liu2018data}
\bibfield{author}{\bibinfo{person}{Zhicheng Liu}, \bibinfo{person}{John Thompson}, \bibinfo{person}{Alan Wilson}, \bibinfo{person}{Mira Dontcheva}, \bibinfo{person}{James Delorey}, \bibinfo{person}{Sam Grigg}, \bibinfo{person}{Bernard Kerr}, {and} \bibinfo{person}{John Stasko}.} \bibinfo{year}{2018}\natexlab{}.
\newblock \showarticletitle{{Data Illustrator}: Augmenting Vector Design Tools with Lazy Data Binding for Expressive Visualization Authoring}. In \bibinfo{booktitle}{\emph{Proceedings of the ACM Conference on Human Factors in Computing Systems (CHI)}}. \bibinfo{pages}{123:1--13}.
\newblock
\urldef\tempurl%
\url{https://doi.org/10.1145/3173574.3173697}
\showDOI{\tempurl}


\bibitem[Lozhkov et~al\mbox{.}(2024)]%
        {lozhkov2024starcoder}
\bibfield{author}{\bibinfo{person}{Anton Lozhkov}, \bibinfo{person}{Raymond Li}, \bibinfo{person}{Loubna~Ben Allal}, \bibinfo{person}{Federico Cassano}, \bibinfo{person}{Joel Lamy-Poirier}, \bibinfo{person}{Nouamane Tazi}, \bibinfo{person}{Ao Tang}, \bibinfo{person}{Dmytro Pykhtar}, \bibinfo{person}{Jiawei Liu}, \bibinfo{person}{Yuxiang Wei}, {et~al\mbox{.}}} \bibinfo{year}{2024}\natexlab{}.
\newblock \showarticletitle{Starcoder 2 and the stack v2: The next generation}.
\newblock \bibinfo{journal}{\emph{arXiv preprint arXiv:2402.19173}} (\bibinfo{year}{2024}).
\newblock


\bibitem[Luo et~al\mbox{.}(2022)]%
        {luo2021natural}
\bibfield{author}{\bibinfo{person}{Yuyu Luo}, \bibinfo{person}{Nan Tang}, \bibinfo{person}{Guoliang Li}, \bibinfo{person}{Jiawei Tang}, \bibinfo{person}{Chengliang Chai}, {and} \bibinfo{person}{Xuedi Qin}.} \bibinfo{year}{2022}\natexlab{}.
\newblock \showarticletitle{Natural Language to Visualization by Neural Machine Translation}.
\newblock \bibinfo{journal}{\emph{{IEEE} Trans. Vis. Comput. Graph.}} \bibinfo{volume}{28}, \bibinfo{number}{1} (\bibinfo{year}{2022}), \bibinfo{pages}{217--226}.
\newblock
\urldef\tempurl%
\url{https://doi.org/10.1109/TVCG.2021.3114848}
\showDOI{\tempurl}


\bibitem[Maddigan and Susnjak(2023)]%
        {maddigan2023chat2vis}
\bibfield{author}{\bibinfo{person}{Paula Maddigan} {and} \bibinfo{person}{Teo Susnjak}.} \bibinfo{year}{2023}\natexlab{}.
\newblock \showarticletitle{Chat2vis: Generating data visualisations via natural language using chatgpt, codex and gpt-3 large language models}.
\newblock \bibinfo{journal}{\emph{Ieee Access}} (\bibinfo{year}{2023}).
\newblock


\bibitem[Masson et~al\mbox{.}(2024)]%
        {masson2024directgpt}
\bibfield{author}{\bibinfo{person}{Damien Masson}, \bibinfo{person}{Sylvain Malacria}, \bibinfo{person}{G{\'e}ry Casiez}, {and} \bibinfo{person}{Daniel Vogel}.} \bibinfo{year}{2024}\natexlab{}.
\newblock \showarticletitle{Directgpt: A direct manipulation interface to interact with large language models}. In \bibinfo{booktitle}{\emph{Proceedings of the CHI Conference on Human Factors in Computing Systems}}. \bibinfo{pages}{1--16}.
\newblock


\bibitem[McNutt et~al\mbox{.}(2023)]%
        {mcnutt2023design}
\bibfield{author}{\bibinfo{person}{Andrew~M McNutt}, \bibinfo{person}{Chenglong Wang}, \bibinfo{person}{Robert~A Deline}, {and} \bibinfo{person}{Steven~M Drucker}.} \bibinfo{year}{2023}\natexlab{}.
\newblock \showarticletitle{On the design of ai-powered code assistants for notebooks}. In \bibinfo{booktitle}{\emph{Proceedings of the 2023 CHI Conference on Human Factors in Computing Systems}}. \bibinfo{pages}{1--16}.
\newblock


\bibitem[Mitra et~al\mbox{.}(2022)]%
        {mitra2022facilitating}
\bibfield{author}{\bibinfo{person}{Rishab Mitra}, \bibinfo{person}{Arpit Narechania}, \bibinfo{person}{Alex Endert}, {and} \bibinfo{person}{John Stasko}.} \bibinfo{year}{2022}\natexlab{}.
\newblock \showarticletitle{Facilitating conversational interaction in natural language interfaces for visualization}. In \bibinfo{booktitle}{\emph{2022 IEEE Visualization and Visual Analytics (VIS)}}. IEEE, \bibinfo{pages}{6--10}.
\newblock


\bibitem[Moritz et~al\mbox{.}(2019)]%
        {moritz2018formalizing}
\bibfield{author}{\bibinfo{person}{Dominik Moritz}, \bibinfo{person}{Chenglong Wang}, \bibinfo{person}{Greg~L. Nelson}, \bibinfo{person}{Halden Lin}, \bibinfo{person}{Adam~M. Smith}, \bibinfo{person}{Bill Howe}, {and} \bibinfo{person}{Jeffrey Heer}.} \bibinfo{year}{2019}\natexlab{}.
\newblock \showarticletitle{Formalizing Visualization Design Knowledge as Constraints: Actionable and Extensible Models in Draco}.
\newblock \bibinfo{journal}{\emph{{IEEE} Trans. Vis. Comput. Graph.}} \bibinfo{volume}{25}, \bibinfo{number}{1} (\bibinfo{year}{2019}), \bibinfo{pages}{438--448}.
\newblock
\urldef\tempurl%
\url{https://doi.org/10.1109/TVCG.2018.2865240}
\showDOI{\tempurl}


\bibitem[Narechania et~al\mbox{.}(2021)]%
        {narechania2020nl4dv}
\bibfield{author}{\bibinfo{person}{Arpit Narechania}, \bibinfo{person}{Arjun Srinivasan}, {and} \bibinfo{person}{John~T. Stasko}.} \bibinfo{year}{2021}\natexlab{}.
\newblock \showarticletitle{{NL4DV:} {A} Toolkit for Generating Analytic Specifications for Data Visualization from Natural Language Queries}.
\newblock \bibinfo{journal}{\emph{{IEEE} Trans. Vis. Comput. Graph.}} \bibinfo{volume}{27}, \bibinfo{number}{2} (\bibinfo{year}{2021}), \bibinfo{pages}{369--379}.
\newblock
\urldef\tempurl%
\url{https://doi.org/10.1109/TVCG.2020.3030378}
\showDOI{\tempurl}


\bibitem[Observable({[n.\,d.]})]%
        {observable}
\bibfield{author}{\bibinfo{person}{Observable}.} \bibinfo{year}{[n.\,d.]}\natexlab{}.
\newblock \bibinfo{booktitle}{\emph{https://observablehq.com/}}.
\newblock


\bibitem[Olausson et~al\mbox{.}(2023)]%
        {olausson2023self}
\bibfield{author}{\bibinfo{person}{Theo~X Olausson}, \bibinfo{person}{Jeevana~Priya Inala}, \bibinfo{person}{Chenglong Wang}, \bibinfo{person}{Jianfeng Gao}, {and} \bibinfo{person}{Armando Solar-Lezama}.} \bibinfo{year}{2023}\natexlab{}.
\newblock \showarticletitle{Is Self-Repair a Silver Bullet for Code Generation?}. In \bibinfo{booktitle}{\emph{The Twelfth International Conference on Learning Representations}}.
\newblock


\bibitem[OpenAI et~al\mbox{.}(2024)]%
        {openai2024gpt4technicalreport}
\bibfield{author}{\bibinfo{person}{OpenAI}, \bibinfo{person}{Josh Achiam}, \bibinfo{person}{Steven Adler}, \bibinfo{person}{Sandhini Agarwal}, \bibinfo{person}{Lama Ahmad}, \bibinfo{person}{Ilge Akkaya}, \bibinfo{person}{Florencia~Leoni Aleman}, \bibinfo{person}{Diogo Almeida}, \bibinfo{person}{Janko Altenschmidt}, {and} \bibinfo{person}{Sam~Altman et al.}} \bibinfo{year}{2024}\natexlab{}.
\newblock \bibinfo{title}{GPT-4 Technical Report}.
\newblock
\newblock
\showeprint[arxiv]{2303.08774}~[cs.CL]
\urldef\tempurl%
\url{https://arxiv.org/abs/2303.08774}
\showURL{%
\tempurl}


\bibitem[pandas~development team(2023)]%
        {the_pandas_development_team_2023_7741580}
\bibfield{author}{\bibinfo{person}{The pandas~development team}.} \bibinfo{year}{2023}\natexlab{}.
\newblock \bibinfo{booktitle}{\emph{pandas-dev/pandas: Pandas}}.
\newblock
\urldef\tempurl%
\url{https://doi.org/10.5281/zenodo.7741580}
\showDOI{\tempurl}


\bibitem[Polozov and Gulwani(2015)]%
        {polozov2015flashmeta}
\bibfield{author}{\bibinfo{person}{Oleksandr Polozov} {and} \bibinfo{person}{Sumit Gulwani}.} \bibinfo{year}{2015}\natexlab{}.
\newblock \showarticletitle{FlashMeta: a framework for inductive program synthesis}. In \bibinfo{booktitle}{\emph{Object-Oriented Programming, Systems, Languages, and Applications, {OOPSLA} 2015, Pittsburgh, PA, USA, October 25-30, 2015}}, \bibfield{editor}{\bibinfo{person}{Jonathan Aldrich} {and} \bibinfo{person}{Patrick Eugster}} (Eds.). \bibinfo{publisher}{{ACM}}, \bibinfo{pages}{107--126}.
\newblock
\urldef\tempurl%
\url{https://doi.org/10.1145/2814270.2814310}
\showDOI{\tempurl}


\bibitem[Raman and Hellerstein(2001)]%
        {raman2001potter}
\bibfield{author}{\bibinfo{person}{Vijayshankar Raman} {and} \bibinfo{person}{Joseph~M. Hellerstein}.} \bibinfo{year}{2001}\natexlab{}.
\newblock \showarticletitle{Potter's Wheel: An Interactive Data Cleaning System}. In \bibinfo{booktitle}{\emph{{VLDB} 2001, Proceedings of 27th International Conference on Very Large Data Bases, September 11-14, 2001, Roma, Italy}}, \bibfield{editor}{\bibinfo{person}{Peter M.~G. Apers}, \bibinfo{person}{Paolo Atzeni}, \bibinfo{person}{Stefano Ceri}, \bibinfo{person}{Stefano Paraboschi}, \bibinfo{person}{Kotagiri Ramamohanarao}, {and} \bibinfo{person}{Richard~T. Snodgrass}} (Eds.). \bibinfo{publisher}{Morgan Kaufmann}, \bibinfo{pages}{381--390}.
\newblock
\urldef\tempurl%
\url{http://www.vldb.org/conf/2001/P381.pdf}
\showURL{%
\tempurl}


\bibitem[Reid et~al\mbox{.}(2024)]%
        {reid2024gemini}
\bibfield{author}{\bibinfo{person}{Machel Reid}, \bibinfo{person}{Nikolay Savinov}, \bibinfo{person}{Denis Teplyashin}, \bibinfo{person}{Dmitry Lepikhin}, \bibinfo{person}{Timothy Lillicrap}, \bibinfo{person}{Jean-baptiste Alayrac}, \bibinfo{person}{Radu Soricut}, \bibinfo{person}{Angeliki Lazaridou}, \bibinfo{person}{Orhan Firat}, \bibinfo{person}{Julian Schrittwieser}, {et~al\mbox{.}}} \bibinfo{year}{2024}\natexlab{}.
\newblock \showarticletitle{Gemini 1.5: Unlocking multimodal understanding across millions of tokens of context}.
\newblock \bibinfo{journal}{\emph{arXiv preprint arXiv:2403.05530}} (\bibinfo{year}{2024}).
\newblock


\bibitem[Ren et~al\mbox{.}(2017)]%
        {ren2017chartaccent}
\bibfield{author}{\bibinfo{person}{Donghao Ren}, \bibinfo{person}{Matthew Brehmer}, \bibinfo{person}{Bongshin Lee}, \bibinfo{person}{Tobias H{\"o}llerer}, {and} \bibinfo{person}{Eun~Kyoung Choe}.} \bibinfo{year}{2017}\natexlab{}.
\newblock \showarticletitle{Chartaccent: Annotation for data-driven storytelling}. In \bibinfo{booktitle}{\emph{2017 IEEE Pacific Visualization Symposium (PacificVis)}}. Ieee, \bibinfo{pages}{230--239}.
\newblock


\bibitem[Ren et~al\mbox{.}(2019)]%
        {ren2019charticulator}
\bibfield{author}{\bibinfo{person}{Donghao Ren}, \bibinfo{person}{Bongshin Lee}, {and} \bibinfo{person}{Matthew Brehmer}.} \bibinfo{year}{2019}\natexlab{}.
\newblock \showarticletitle{Charticulator: Interactive Construction of Bespoke Chart Layouts}.
\newblock \bibinfo{journal}{\emph{{IEEE} Trans. Vis. Comput. Graph. (Proceedings of InfoVis)}} \bibinfo{volume}{25}, \bibinfo{number}{1} (\bibinfo{year}{2019}).
\newblock
\urldef\tempurl%
\url{https://doi.org/10.1109/TVCG.2018.2865158}
\showDOI{\tempurl}


\bibitem[Ren et~al\mbox{.}(2018)]%
        {ren2018reflecting}
\bibfield{author}{\bibinfo{person}{Donghao Ren}, \bibinfo{person}{Bongshin Lee}, \bibinfo{person}{Matthew Brehmer}, {and} \bibinfo{person}{Nathalie~Henry Riche}.} \bibinfo{year}{2018}\natexlab{}.
\newblock \showarticletitle{Reflecting on the Evaluation of Visualization Authoring Systems : Position Paper}. In \bibinfo{booktitle}{\emph{2018 {IEEE} Evaluation and Beyond - Methodological Approaches for Visualization, {BELIV} 2018, Berlin, Germany, October 21, 2018}}, \bibfield{editor}{\bibinfo{person}{Michael Sedlmair}, \bibinfo{person}{Petra Isenberg}, \bibinfo{person}{Miriah Meyer}, {and} \bibinfo{person}{Tobias Isenberg}} (Eds.). \bibinfo{publisher}{{IEEE} Computer Society}, \bibinfo{pages}{86--92}.
\newblock
\urldef\tempurl%
\url{https://doi.org/10.1109/BELIV.2018.8634297}
\showDOI{\tempurl}


\bibitem[Rule et~al\mbox{.}(2018)]%
        {rule2018exploration}
\bibfield{author}{\bibinfo{person}{Adam Rule}, \bibinfo{person}{Aur{\'e}lien Tabard}, {and} \bibinfo{person}{James~D Hollan}.} \bibinfo{year}{2018}\natexlab{}.
\newblock \showarticletitle{Exploration and explanation in computational notebooks}. In \bibinfo{booktitle}{\emph{Proceedings of the 2018 CHI Conference on Human Factors in Computing Systems}}. \bibinfo{pages}{1--12}.
\newblock


\bibitem[Satyanarayan and Heer(2014)]%
        {satyanarayan2014lyra}
\bibfield{author}{\bibinfo{person}{A. Satyanarayan} {and} \bibinfo{person}{J. Heer}.} \bibinfo{year}{2014}\natexlab{}.
\newblock \showarticletitle{{Lyra: An interactive visualization design environment}}.
\newblock \bibinfo{journal}{\emph{Computer Graphics Forum (Proceedings of EuroVis)}} \bibinfo{volume}{33}, \bibinfo{number}{3} (\bibinfo{year}{2014}).
\newblock
\urldef\tempurl%
\url{https://doi.org/10.1111/cgf.12391}
\showDOI{\tempurl}


\bibitem[Satyanarayan et~al\mbox{.}(2017)]%
        {satyanarayan2017vegalite}
\bibfield{author}{\bibinfo{person}{Arvind Satyanarayan}, \bibinfo{person}{Dominik Moritz}, \bibinfo{person}{Kanit Wongsuphasawat}, {and} \bibinfo{person}{Jeffrey Heer}.} \bibinfo{year}{2017}\natexlab{}.
\newblock \showarticletitle{{Vega-Lite}: A grammar of interactive graphics}.
\newblock \bibinfo{journal}{\emph{IEEE Transactions on Visualization and Computer Graphics (Proceedings of InfoVis)}} \bibinfo{volume}{23}, \bibinfo{number}{1} (\bibinfo{year}{2017}), \bibinfo{pages}{341--350}.
\newblock
\urldef\tempurl%
\url{https://doi.org/10.1109/TVCG.2016.2599030}
\showDOI{\tempurl}


\bibitem[Shi et~al\mbox{.}(2019)]%
        {shi2019task}
\bibfield{author}{\bibinfo{person}{Danqing Shi}, \bibinfo{person}{Yang Shi}, \bibinfo{person}{Xinyue Xu}, \bibinfo{person}{Nan Chen}, \bibinfo{person}{Siwei Fu}, \bibinfo{person}{Hongjin Wu}, {and} \bibinfo{person}{Nan Cao}.} \bibinfo{year}{2019}\natexlab{}.
\newblock \showarticletitle{Task-oriented optimal sequencing of visualization charts}. In \bibinfo{booktitle}{\emph{2019 IEEE Visualization in Data Science (VDS)}}. IEEE, \bibinfo{pages}{58--66}.
\newblock


\bibitem[Stolte et~al\mbox{.}(2002)]%
        {stolte2002query}
\bibfield{author}{\bibinfo{person}{Chris Stolte}, \bibinfo{person}{Diane Tang}, {and} \bibinfo{person}{Pat Hanrahan}.} \bibinfo{year}{2002}\natexlab{}.
\newblock \showarticletitle{Query, analysis, and visualization of hierarchically structured data using Polaris}. In \bibinfo{booktitle}{\emph{Proceedings of the Eighth {ACM} {SIGKDD} International Conference on Knowledge Discovery and Data Mining, July 23-26, 2002, Edmonton, Alberta, Canada}}. \bibinfo{publisher}{{ACM}}, \bibinfo{pages}{112--122}.
\newblock
\urldef\tempurl%
\url{https://doi.org/10.1145/775047.775064}
\showDOI{\tempurl}


\bibitem[Tankelevitch et~al\mbox{.}(2024)]%
        {DBLP:conf/chi/TankelevitchKSS24}
\bibfield{author}{\bibinfo{person}{Lev Tankelevitch}, \bibinfo{person}{Viktor Kewenig}, \bibinfo{person}{Auste Simkute}, \bibinfo{person}{Ava~Elizabeth Scott}, \bibinfo{person}{Advait Sarkar}, \bibinfo{person}{Abigail Sellen}, {and} \bibinfo{person}{Sean Rintel}.} \bibinfo{year}{2024}\natexlab{}.
\newblock \showarticletitle{The Metacognitive Demands and Opportunities of Generative {AI}}. In \bibinfo{booktitle}{\emph{Proceedings of the {CHI} Conference on Human Factors in Computing Systems, {CHI} 2024, Honolulu, HI, USA, May 11-16, 2024}}, \bibfield{editor}{\bibinfo{person}{Florian~'Floyd' Mueller}, \bibinfo{person}{Penny Kyburz}, \bibinfo{person}{Julie~R. Williamson}, \bibinfo{person}{Corina Sas}, \bibinfo{person}{Max~L. Wilson}, \bibinfo{person}{Phoebe O.~Toups Dugas}, {and} \bibinfo{person}{Irina Shklovski}} (Eds.). \bibinfo{publisher}{{ACM}}, \bibinfo{pages}{680:1--680:24}.
\newblock
\urldef\tempurl%
\url{https://doi.org/10.1145/3613904.3642902}
\showDOI{\tempurl}


\bibitem[Tian et~al\mbox{.}(2024)]%
        {tian2024chartgpt}
\bibfield{author}{\bibinfo{person}{Yuan Tian}, \bibinfo{person}{Weiwei Cui}, \bibinfo{person}{Dazhen Deng}, \bibinfo{person}{Xinjing Yi}, \bibinfo{person}{Yurun Yang}, \bibinfo{person}{Haidong Zhang}, {and} \bibinfo{person}{Yingcai Wu}.} \bibinfo{year}{2024}\natexlab{}.
\newblock \showarticletitle{Chartgpt: Leveraging llms to generate charts from abstract natural language}.
\newblock \bibinfo{journal}{\emph{IEEE Transactions on Visualization and Computer Graphics}} (\bibinfo{year}{2024}).
\newblock


\bibitem[Touvron et~al\mbox{.}(2023)]%
        {touvron2023llama}
\bibfield{author}{\bibinfo{person}{Hugo Touvron}, \bibinfo{person}{Thibaut Lavril}, \bibinfo{person}{Gautier Izacard}, \bibinfo{person}{Xavier Martinet}, \bibinfo{person}{Marie-Anne Lachaux}, \bibinfo{person}{Timoth{\'e}e Lacroix}, \bibinfo{person}{Baptiste Rozi{\`e}re}, \bibinfo{person}{Naman Goyal}, \bibinfo{person}{Eric Hambro}, \bibinfo{person}{Faisal Azhar}, {et~al\mbox{.}}} \bibinfo{year}{2023}\natexlab{}.
\newblock \showarticletitle{Llama: Open and efficient foundation language models}.
\newblock \bibinfo{journal}{\emph{arXiv preprint arXiv:2302.13971}} (\bibinfo{year}{2023}).
\newblock


\bibitem[Vaithilingam et~al\mbox{.}(2024)]%
        {vaithilingam2024dynavis}
\bibfield{author}{\bibinfo{person}{Priyan Vaithilingam}, \bibinfo{person}{Elena~L Glassman}, \bibinfo{person}{Jeevana~Priya Inala}, {and} \bibinfo{person}{Chenglong Wang}.} \bibinfo{year}{2024}\natexlab{}.
\newblock \showarticletitle{DynaVis: Dynamically Synthesized UI Widgets for Visualization Editing}. In \bibinfo{booktitle}{\emph{Proceedings of the CHI Conference on Human Factors in Computing Systems}}. \bibinfo{pages}{1--17}.
\newblock


\bibitem[VanderPlas et~al\mbox{.}(2018)]%
        {vanderplas2018altair}
\bibfield{author}{\bibinfo{person}{Jacob VanderPlas}, \bibinfo{person}{Brian~E. Granger}, \bibinfo{person}{Jeffrey Heer}, \bibinfo{person}{Dominik Moritz}, \bibinfo{person}{Kanit Wongsuphasawat}, \bibinfo{person}{Arvind Satyanarayan}, \bibinfo{person}{Eitan Lees}, \bibinfo{person}{Ilia Timofeev}, \bibinfo{person}{Ben Welsh}, {and} \bibinfo{person}{Scott Sievert}.} \bibinfo{year}{2018}\natexlab{}.
\newblock \showarticletitle{Altair: Interactive Statistical Visualizations for Python}.
\newblock \bibinfo{journal}{\emph{J. Open Source Softw.}} \bibinfo{volume}{3}, \bibinfo{number}{32}, \bibinfo{pages}{1057}.
\newblock
\urldef\tempurl%
\url{https://doi.org/10.21105/joss.01057}
\showDOI{\tempurl}


\bibitem[Wang et~al\mbox{.}(2021)]%
        {wang2021falx}
\bibfield{author}{\bibinfo{person}{Chenglong Wang}, \bibinfo{person}{Yu Feng}, \bibinfo{person}{Rastislav Bod{\'{\i}}k}, \bibinfo{person}{Isil Dillig}, \bibinfo{person}{Alvin Cheung}, {and} \bibinfo{person}{Amy~J. Ko}.} \bibinfo{year}{2021}\natexlab{}.
\newblock \showarticletitle{Falx: Synthesis-Powered Visualization Authoring}. In \bibinfo{booktitle}{\emph{{CHI} '21: {CHI} Conference on Human Factors in Computing Systems, Virtual Event / Yokohama, Japan, May 8-13, 2021}}, \bibfield{editor}{\bibinfo{person}{Yoshifumi Kitamura}, \bibinfo{person}{Aaron Quigley}, \bibinfo{person}{Katherine Isbister}, \bibinfo{person}{Takeo Igarashi}, \bibinfo{person}{Pernille Bj{\o}rn}, {and} \bibinfo{person}{Steven~Mark Drucker}} (Eds.). \bibinfo{publisher}{{ACM}}, \bibinfo{pages}{106:1--106:15}.
\newblock
\urldef\tempurl%
\url{https://doi.org/10.1145/3411764.3445249}
\showDOI{\tempurl}


\bibitem[Wang et~al\mbox{.}(2023)]%
        {wang2023data}
\bibfield{author}{\bibinfo{person}{Chenglong Wang}, \bibinfo{person}{John Thompson}, {and} \bibinfo{person}{Bongshin Lee}.} \bibinfo{year}{2023}\natexlab{}.
\newblock \showarticletitle{Data Formulator: Ai-powered concept-driven visualization authoring}.
\newblock \bibinfo{journal}{\emph{IEEE Transactions on Visualization and Computer Graphics}} (\bibinfo{year}{2023}).
\newblock


\bibitem[Wei et~al\mbox{.}(2022)]%
        {wei2022chain}
\bibfield{author}{\bibinfo{person}{Jason Wei}, \bibinfo{person}{Xuezhi Wang}, \bibinfo{person}{Dale Schuurmans}, \bibinfo{person}{Maarten Bosma}, \bibinfo{person}{Fei Xia}, \bibinfo{person}{Ed Chi}, \bibinfo{person}{Quoc~V Le}, \bibinfo{person}{Denny Zhou}, {et~al\mbox{.}}} \bibinfo{year}{2022}\natexlab{}.
\newblock \showarticletitle{Chain-of-thought prompting elicits reasoning in large language models}.
\newblock \bibinfo{journal}{\emph{Advances in neural information processing systems}}  \bibinfo{volume}{35} (\bibinfo{year}{2022}), \bibinfo{pages}{24824--24837}.
\newblock


\bibitem[White and Drucker(2007)]%
        {white2007investigating}
\bibfield{author}{\bibinfo{person}{Ryen~W White} {and} \bibinfo{person}{Steven~M Drucker}.} \bibinfo{year}{2007}\natexlab{}.
\newblock \showarticletitle{Investigating behavioral variability in web search}. In \bibinfo{booktitle}{\emph{Proceedings of the 16th international conference on World Wide Web}}. \bibinfo{pages}{21--30}.
\newblock


\bibitem[Wickham(2009)]%
        {wickham2009ggplot2}
\bibfield{author}{\bibinfo{person}{Hadley Wickham}.} \bibinfo{year}{2009}\natexlab{}.
\newblock \bibinfo{booktitle}{\emph{ggplot2 - Elegant Graphics for Data Analysis}}.
\newblock \bibinfo{publisher}{Springer}.
\newblock
\showISBNx{978-0-387-98140-6}
\urldef\tempurl%
\url{https://doi.org/10.1007/978-0-387-98141-3}
\showDOI{\tempurl}


\bibitem[Wickham(2014)]%
        {wickham2014tidy-data}
\bibfield{author}{\bibinfo{person}{Hadley Wickham}.} \bibinfo{year}{2014}\natexlab{}.
\newblock \showarticletitle{Tidy data}.
\newblock \bibinfo{journal}{\emph{The Journal of Statistical Software}}  \bibinfo{volume}{59} (\bibinfo{year}{2014}).
\newblock
Issue 10.
\urldef\tempurl%
\url{http://www.jstatsoft.org/v59/i10/}
\showURL{%
\tempurl}


\bibitem[Wickham et~al\mbox{.}(2019)]%
        {wickham2019tidyverse}
\bibfield{author}{\bibinfo{person}{Hadley Wickham}, \bibinfo{person}{Mara Averick}, \bibinfo{person}{Jennifer Bryan}, \bibinfo{person}{Winston Chang}, \bibinfo{person}{Lucy McGowan}, \bibinfo{person}{Romain Fran{\c c}ois}, \bibinfo{person}{Garrett Grolemund}, \bibinfo{person}{Alex Hayes}, \bibinfo{person}{Lionel Henry}, \bibinfo{person}{Jim Hester}, \bibinfo{person}{Max Kuhn}, \bibinfo{person}{Thomas Pedersen}, \bibinfo{person}{Evan Miller}, \bibinfo{person}{Stephan Bache}, \bibinfo{person}{Kirill M{\"u}ller}, \bibinfo{person}{Jeroen Ooms}, \bibinfo{person}{David Robinson}, \bibinfo{person}{Dana Seidel}, \bibinfo{person}{Vitalie Spinu}, \bibinfo{person}{Kohske Takahashi}, \bibinfo{person}{Davis Vaughan}, \bibinfo{person}{Claus Wilke}, \bibinfo{person}{Kara Woo}, {and} \bibinfo{person}{Hiroaki Yutani}.} \bibinfo{year}{2019}\natexlab{}.
\newblock \showarticletitle{Welcome to the tidyverse}.
\newblock \bibinfo{journal}{\emph{J. Open Source Softw.}} \bibinfo{volume}{4}, \bibinfo{number}{43} (\bibinfo{date}{Nov.} \bibinfo{year}{2019}), \bibinfo{pages}{1686}.
\newblock
\showISSN{2475-9066}
\urldef\tempurl%
\url{https://doi.org/10.21105/joss.01686}
\showDOI{\tempurl}


\bibitem[Wilkinson(2005)]%
        {DBLP:books/daglib/0024564}
\bibfield{author}{\bibinfo{person}{Leland Wilkinson}.} \bibinfo{year}{2005}\natexlab{}.
\newblock \bibinfo{booktitle}{\emph{The Grammar of Graphics, Second Edition}}.
\newblock \bibinfo{publisher}{Springer}.
\newblock
\showISBNx{978-0-387-24544-7}


\bibitem[Wongsuphasawat et~al\mbox{.}(2017)]%
        {Wongsuphasawat2017voyager2}
\bibfield{author}{\bibinfo{person}{Kanit Wongsuphasawat}, \bibinfo{person}{Zening Qu}, \bibinfo{person}{Dominik Moritz}, \bibinfo{person}{Riley Chang}, \bibinfo{person}{Felix Ouk}, \bibinfo{person}{Anushka Anand}, \bibinfo{person}{Jock Mackinlay}, \bibinfo{person}{Bill Howe}, {and} \bibinfo{person}{Jeffrey Heer}.} \bibinfo{year}{2017}\natexlab{}.
\newblock \showarticletitle{Voyager 2: Augmenting Visual Analysis with Partial View Specifications}. In \bibinfo{booktitle}{\emph{Proceedings of the 2017 CHI Conference on Human Factors in Computing Systems}} (Denver, Colorado, USA) \emph{(\bibinfo{series}{CHI '17})}. \bibinfo{publisher}{Association for Computing Machinery}, \bibinfo{address}{New York, NY, USA}, \bibinfo{pages}{2648–2659}.
\newblock
\showISBNx{9781450346559}
\urldef\tempurl%
\url{https://doi.org/10.1145/3025453.3025768}
\showDOI{\tempurl}


\bibitem[Wu et~al\mbox{.}(2023)]%
        {wu2023autogen}
\bibfield{author}{\bibinfo{person}{Qingyun Wu}, \bibinfo{person}{Gagan Bansal}, \bibinfo{person}{Jieyu Zhang}, \bibinfo{person}{Yiran Wu}, \bibinfo{person}{Shaokun Zhang}, \bibinfo{person}{Erkang Zhu}, \bibinfo{person}{Beibin Li}, \bibinfo{person}{Li Jiang}, \bibinfo{person}{Xiaoyun Zhang}, {and} \bibinfo{person}{Chi Wang}.} \bibinfo{year}{2023}\natexlab{}.
\newblock \showarticletitle{Autogen: Enabling next-gen llm applications via multi-agent conversation framework}.
\newblock \bibinfo{journal}{\emph{arXiv preprint arXiv:2308.08155}} (\bibinfo{year}{2023}).
\newblock


\bibitem[Wu et~al\mbox{.}(2022)]%
        {wu2022nl2viz}
\bibfield{author}{\bibinfo{person}{Zhengkai Wu}, \bibinfo{person}{Vu Le}, \bibinfo{person}{Ashish Tiwari}, \bibinfo{person}{Sumit Gulwani}, \bibinfo{person}{Arjun Radhakrishna}, \bibinfo{person}{Ivan Radi{\v{c}}ek}, \bibinfo{person}{Gustavo Soares}, \bibinfo{person}{Xinyu Wang}, \bibinfo{person}{Zhenwen Li}, {and} \bibinfo{person}{Tao Xie}.} \bibinfo{year}{2022}\natexlab{}.
\newblock \showarticletitle{NL2Viz: natural language to visualization via constrained syntax-guided synthesis}. In \bibinfo{booktitle}{\emph{Proceedings of the 30th ACM Joint European Software Engineering Conference and Symposium on the Foundations of Software Engineering}}. \bibinfo{pages}{972--983}.
\newblock


\bibitem[Xiong et~al\mbox{.}(2022)]%
        {xiong2022visualizing}
\bibfield{author}{\bibinfo{person}{Kai Xiong}, \bibinfo{person}{Siwei Fu}, \bibinfo{person}{Guoming Ding}, \bibinfo{person}{Zhongsu Luo}, \bibinfo{person}{Rong Yu}, \bibinfo{person}{Wei Chen}, \bibinfo{person}{Hujun Bao}, {and} \bibinfo{person}{Yingcai Wu}.} \bibinfo{year}{2022}\natexlab{}.
\newblock \showarticletitle{Visualizing the scripts of data wrangling with SOMNUS}.
\newblock \bibinfo{journal}{\emph{IEEE Transactions on Visualization and Computer Graphics}} (\bibinfo{year}{2022}).
\newblock


\bibitem[Zamfirescu-Pereira et~al\mbox{.}(2023)]%
        {zamfirescu2023johnny}
\bibfield{author}{\bibinfo{person}{JD Zamfirescu-Pereira}, \bibinfo{person}{Richmond~Y Wong}, \bibinfo{person}{Bjoern Hartmann}, {and} \bibinfo{person}{Qian Yang}.} \bibinfo{year}{2023}\natexlab{}.
\newblock \showarticletitle{Why Johnny can’t prompt: how non-AI experts try (and fail) to design LLM prompts}. In \bibinfo{booktitle}{\emph{Proceedings of the 2023 CHI Conference on Human Factors in Computing Systems}}. \bibinfo{pages}{1--21}.
\newblock


\bibitem[Zhang et~al\mbox{.}(2023)]%
        {zhang2023tell}
\bibfield{author}{\bibinfo{person}{Qingru Zhang}, \bibinfo{person}{Chandan Singh}, \bibinfo{person}{Liyuan Liu}, \bibinfo{person}{Xiaodong Liu}, \bibinfo{person}{Bin Yu}, \bibinfo{person}{Jianfeng Gao}, {and} \bibinfo{person}{Tuo Zhao}.} \bibinfo{year}{2023}\natexlab{}.
\newblock \showarticletitle{Tell your model where to attend: Post-hoc attention steering for llms}.
\newblock \bibinfo{journal}{\emph{arXiv preprint arXiv:2311.02262}} (\bibinfo{year}{2023}).
\newblock


\bibitem[Zhang et~al\mbox{.}(2024)]%
        {zhang2024training}
\bibfield{author}{\bibinfo{person}{Shaokun Zhang}, \bibinfo{person}{Jieyu Zhang}, \bibinfo{person}{Jiale Liu}, \bibinfo{person}{Linxin Song}, \bibinfo{person}{Chi Wang}, \bibinfo{person}{Ranjay Krishna}, {and} \bibinfo{person}{Qingyun Wu}.} \bibinfo{year}{2024}\natexlab{}.
\newblock \showarticletitle{Training Language Model Agents without Modifying Language Models}.
\newblock \bibinfo{journal}{\emph{ICML'24}} (\bibinfo{year}{2024}).
\newblock


\bibitem[Zheng et~al\mbox{.}(2024)]%
        {zheng2024opencodeinterpreter}
\bibfield{author}{\bibinfo{person}{Tianyu Zheng}, \bibinfo{person}{Ge Zhang}, \bibinfo{person}{Tianhao Shen}, \bibinfo{person}{Xueling Liu}, \bibinfo{person}{Bill~Yuchen Lin}, \bibinfo{person}{Jie Fu}, \bibinfo{person}{Wenhu Chen}, {and} \bibinfo{person}{Xiang Yue}.} \bibinfo{year}{2024}\natexlab{}.
\newblock \showarticletitle{OpenCodeInterpreter: Integrating Code Generation with Execution and Refinement}.
\newblock \bibinfo{journal}{\emph{arXiv preprint arXiv:2402.14658}} (\bibinfo{year}{2024}).
\newblock


\end{thebibliography}



\end{document}